\documentclass[pra,preprint,nobalancelastpage,onecolumn,nofootinbib,superscriptaddress]{revtex4-2}
\pdfoutput=1
\usepackage[utf8]{inputenc}
\usepackage{color,amsthm,amsmath,amsxtra,amsfonts,dsfont,graphicx,bm,amssymb}
\usepackage{qcircuit}
\usepackage{booktabs}
\usepackage{verbatim}
\usepackage{hyperref}
\usepackage[english]{babel}
\usepackage[dvipsnames]{xcolor}
\usepackage{mathtools}
\hypersetup{hidelinks}
\usepackage{tikz}
\usetikzlibrary{arrows.meta,calc,fit,positioning}

\tikzset{
  graphvertex/.style={
    circle,draw,thick,minimum size=6mm,inner sep=0pt
  },
  endpoint/.style={
    graphvertex,fill=blue!15,draw=blue!65!black
  },
  localtensor/.style={
    rectangle,draw,thick,rounded corners=1pt,
    minimum width=7mm,minimum height=7mm,fill=gray!8
  },
  cedge/.style={thick},
  treeedge/.style={very thick,blue!70!black},
  residual/.style={thick,dashed,red!75!black},
  prune/.style={-{Latex[length=2.4mm]},very thick}
}
\usepackage{graphicx}
\usetikzlibrary{shapes, arrows, positioning, calc}
\usepackage{braket}
\usepackage{float}
\usepackage{subcaption} % for subfigures
\usepackage{empheq}
\newcommand{\tr}[1]{{\mathrm{Tr}\left[#1\right]}}

\newcommand{\di}{\mathrm{d}}
\newtheorem{theorem}{Theorem}
\newtheorem{corollary}{Corollary}

\newcommand{\op}{\mathrm{op}}
\newcommand{\ii}{i}
\newcommand{\Tr}{\operatorname{Tr}}

\newcommand{\norm}[1]{\left\lVert #1\right\rVert}
\newcommand{\abs}[1]{\left\lvert #1\right\rvert}

\usepackage{etoolbox}

\makeatletter
\patchcmd{\frontmatter@abstract@produce}
  {\preprintsty@sw}
  {\@secondoftwo}
  {}
  {\PackageWarning{abstract-placement}{Patch failed}}
\makeatother
\begin{document}
\title{Probing equilibrium properties with Gaussian states beyond mean-field}
\author{Esther Cruz}
\affiliation{Max-Planck-Institut für Quantenoptik, Hans-Kopfermann-Str.~1, D-85748 Garching, Germany}
\affiliation{Munich Center for Quantum Science and Technology (MCQST), Schellingstr.~4, D-80799 München, Germany}
\author{J. Ignacio Cirac} 
\affiliation{Max-Planck-Institut für Quantenoptik, Hans-Kopfermann-Str.~1, D-85748 Garching, Germany}
\affiliation{Munich Center for Quantum Science and Technology (MCQST), Schellingstr.~4, D-80799 München, Germany}
\author{Dominik S. Wild}
\affiliation{Google Quantum AI, D-80636 Munich, Germany}
\begin{abstract}
Computing equilibrium properties of interacting fermionic systems is central to many problems in condensed matter physics, yet computationally challenging. Mean-field methods tackle this problem by restricting to a tractable class of states, which, however, often fail to capture genuine many-body correlations. We illustrate this in the context of Gaussian mean-field methods by minimizing the energy variance of Gaussian states, showing that even states with the narrowest energy distributions poorly reproduce the microcanonical ensemble. To overcome this limitation, we develop a computationally efficient method to construct linear superpositions of Gaussian states whose energy variance can be systematically decreased by adding more terms to the superposition. We provide complementary results for the canonical ensemble, for which we propose a sampling algorithm based on convex combinations of Gaussian states. We benchmark our methods using the $t$--$V$ model and find that both methods greatly outperform mean-field calculations. In particular, we obtain the correct critical exponents as well as a significantly improved estimate for the critical temperature. As an independent result, we prove a quantum Berry--Esseen theorem for fermionic Gaussian states, which states that the local density of states converges to a Gaussian distribution in the thermodynamic limit without any assumptions about the spatial correlations of the state. 
\end{abstract}
\maketitle

\section{Introduction}
The computation of equilibrium properties of interacting fermionic systems is a central problem in theoretical condensed matter physics with far-reaching impact. The ability to rapidly compute local observables and correlation functions would further our understanding of correlated phases of matter and would serve as a powerful tool in the development of new materials. Given the significance of this problem, it has been approached with a wide range of numerical techniques. For example, auxiliary-field quantum Monte Carlo (QMC)~\cite{blankenbecler_monte_1981} enables large-scale studies but it is hampered by the fermionic sign problem at low temperatures and away from half filling~\cite{loh_sign_1990,iglovikov_geometry_2015}. These problems can be avoided using tensor network methods such as DMRG~\cite{white_density_1992} and iPEPS~\cite{jordan_classical_2008, corboz_simulation_2010}, which, however, are typically limited to ground states or low temperature. Even though finite temperature methods exist~\cite{white_minimally_2009,sinha_finite_2022}, their convergence to the thermodynamic limit is often computationally demanding, especially in two dimensions.

Another prominent family of methods is based on fermionic Gaussian states. They are attractive from a computational perspective because Gaussian states are completely characterized by their covariance matrix, and Wick's theorem reduces higher-order correlation functions to two-point functions~\cite{bravyi_lagrangian_2004}. A paradigmatic example is the generalized Hartree--Fock method, which may be viewed as variational optimization over Gaussian states~\cite{kraus_generalized_2010}. Because Gaussian states describe non-interacting fermionic systems exactly, this approach typically works well when interactions are sufficiently weak, or when the relevant physics is captured by an effective mean-field description. By contrast, Gaussian methods typically fail in strongly correlated systems. These limitations have in some cases been overcome using non-Gaussian transformations for which observables remain efficiently computable~\cite{shi_variational_2018}. Alternatively, linear combinations of Gaussian states can enlarge the variational manifold~\cite{fukutome_theory_1988, ikawa_orbital_1993, koch_linear_1993, jimenez-hoyos_multi-component_2013, snyman_efficient_2021, boutin_quantum_2021, giuliani_precise_2026}. When the Gaussian states in the superposition are not fixed, the optimization problem typically becomes highly nonconvex, limiting the method to a small number of Gaussian states in the superposition.

In this work, we use linear combinations of a fixed set of physically inspired Gaussian states to approximate thermal ensembles. We evaluate our methods using the $t$--$V$ model, introduced in section~\ref{section-gaussian}. We compare our results for small systems to exact diagonalization and carry out large-scale computations with up to 324 sites, which we benchmark using previous QMC calculations. We first show in section~\ref{sec:microcanonical} that single Gaussian states fail to adequately describe the microcanonical ensemble. To this end, we develop a constrained optimization, which enables us to minimize the energy variance of a Gaussian state at a fixed mean. The optimization is widely applicable to other observables and may be of interest in its own right. We also prove in Appendix~\ref{appendix-ldos-proof} that for local Hamiltonians with quartic interactions, the local density of states of fermionic Gaussian states approaches a Gaussian distribution in the thermodynamic limit, provided that their energy variance grows at least linearly with system size, an assumption justified by our numerical results for the $t$--$V$ model. To overcome these limitations, we use the time-dependent variational principle (TDVP) to generate Gaussian states at a fixed mean energy. By minimizing the energy variance in the subspace spanned by these states, we are able to construct states that faithfully reproduce the properties of both local observables and the entanglement entropy of energy eigenstates.  We address similar questions concerning the canonical ensemble in section~\ref{sec:canonical}. By sampling over states obtained from TDVP in imaginary time, we are able to reproduce the thermal phase transition of the model in good agreement with QMC calculations. Individual Gaussian states, as obtained by minimizing the free energy, again perform poorly. Beyond estimating the critical temperature more accurately, our method produces the correct critical exponents associated with the universality class of the $t$-$V$ model, which differ from the mean-field critical exponents obtained with single Gaussian states. To conclude, we show in section~\ref{sec:discussion} that the real-time and imaginary-time methods from the preceding sections can be combined to systematically improve the results for the canonical ensemble.

%%%%%%%%%%%%%%%%%%%%%%%%%%%%%%%%%%%%%%%%%%%%%%%%%%%%%%%%%%%%%%%%%%%%%%%%
\section{Setup and notation} \label{section-gaussian}
We consider a set of $N$ fermionic modes described by the creation and annihilation operators $a_n^\dagger$ and $a_n$, respectively. It is convenient to work in the Hermitian Majorana basis $c_n = a_n + a_n^\dagger$ and $c_{N + n} = i (a_n - a_n^\dagger)$, which satisfy the commutation relations $\{ c_m, c_n \} = 2 \delta_{mn}$. In physical terms, fermionic Gaussian states may be viewed as the set of Gibbs states of noninteracting Hamiltonians. A Gaussian state $\rho$ is fully described by the covariance matrix, which is the real, skew-symmetric $2 N \times 2 N$ matrix $\Gamma$ with matrix elements
\begin{equation} \label{eq:covariance-matrix}
    \Gamma_{mn} = \frac{i}{2} \mathrm{Tr} \left( \left[ c_m, c_n \right] \rho \right).
\end{equation}
According to Wick's theorem, any higher-order correlator can be decomposed into a sum of products of two-point correlators.

Gaussian states are described by $O(N^2)$ real parameters, rendering them suitable for efficient classical simulation. An important concept in this context is the time-dependent variational principle (TDVP), which is a locally optimal approximation of non-Gaussian dynamics within the manifold of Gaussian states.
% Let $\mathcal{M}_G$ denote the manifold of fermionic Gaussian states. 
Consider an interacting Hamiltonian of the form
\begin{equation}
    H
    = i\sum_{m,n}T_{mn}c_m c_n
    + \sum_{m,n,p,q}U_{mnpq}c_m c_n c_p c_q ,
\end{equation}
where $T$ can be chosen to be a real, skew-symmetric matrix encoding the non-interacting part of the Hamiltonian and $U$ is a real, fully antisymmetric 4-index tensor describing two-body interactions. When $U \neq 0$, the evolution generated by $H$ generates non-Gaussian correlations. Starting from a Gaussian state $\ket{\psi_G}$, TDVP replaces the exact time derivative $-i H\ket{\psi_G}$ by its projection onto the tangent space of the variational manifold of Gaussian states.

The resulting equation of motion for the covariance matrix is given by
\begin{align}
    \label{eq:real-time-tdvp}
    \dot{\Gamma} =4 \bigl [h(\Gamma),\Gamma\bigr], \qquad
    h(\Gamma) = T + 6 \operatorname{tr}_2(U\Gamma),
\end{align}
where we have defined $\bigl[\operatorname{tr}_2(U\Gamma)\bigr]_{mn}  = \sum_{p,q}U_{mnpq}\Gamma_{qp}$. Thus the TDVP evolution is equivalent to an evolution under a quadratic but state-dependent Hamiltonian $H_{\mathrm{eff}}(\Gamma) = i\sum_{m, n}h_{mn}(\Gamma)c_m c_n$.

TDVP can be similarly applied to imaginary-time evolution, where the state at imaginary time $\tau$ is given by the state $\ket{\psi(\tau)} = e^{- \tau H} \ket{\psi(0)}$, up to normalization, yielding the equation of motion
\begin{equation}
    \dot{\Gamma} =  -4 \Gamma h(\Gamma)\Gamma - 4 h(\Gamma) .
    \label{eq:imag-time-tdvp}
\end{equation}
For a pure initial Gaussian state, this evolution preserves purity, equivalent to the condition $\Gamma^2=-\mathds{1}$, and monotonically decreases the mean energy.

Imaginary-time evolution therefore provides a practical method for obtaining a Gaussian approximation to the ground state~\cite{kraus_generalized_2010,kraus_thesis_fermions}.

To study the performance of our methods, we consider the $t$--$V$ model on a two-dimensional square lattice with periodic boundary conditions. Labeling each site of the lattice by an index $i$, the Hamiltonian may be written as
\begin{equation}\label{eq:tV-model}
	H = J \sum_{\langle i,j\rangle} \left( a_{i}^\dagger a_{j} + \mathrm{h.c.} \right) + V \sum_{\langle i,j\rangle} \left( n_{i} - \frac{1}{2}\right) \left(n_{j}- \frac{1}{2}\right),
\end{equation}
where the sums run over nearest-neighbor pairs and $n_i = a_i^\dagger a_i$. The first term is a hopping term, for which we use the coefficient $J$ instead of namesake $t$ of the model as we prefer to reserve $t$ for time variables. The second term is a nearest-neighbor density-density interaction. One may further include a chemical potential, although we focus here on the point at which the model is symmetric under the particle-hole symmetry $a_i \leftrightarrow a_i^\dagger$. 

The $t$--$V$ model on the square lattice undergoes a phase transition as a function of the temperature for any positive $V$, from an ordered phase at low temperature to a disordered phase at high temperature~\cite{gubernatis_two-dimensional_1985}. Owing to the repulsive nearest-neighbor interaction, the ordered phase breaks the translational symmetry of the lattice, leading to a checkerboard pattern in the density. Following previous works~\cite{gubernatis_two-dimensional_1985,wang_fermionic_2014, wang_stochastic_2016}, we use the order parameter
\begin{equation}\label{eq:staggered-density}
    M_2 = \frac{1}{N^2} \left\langle \left( \sum_i \eta_i n_i \right)^2 \right\rangle
\end{equation}
to diagnose the phase, where $\eta_i = 1$ if $i \in A$ and $\eta_i = -1$ if $i \in B$ with $A$ and $B$ bipartitions of the lattice.

We will benchmark our methods against QMC results~\cite{gubernatis_two-dimensional_1985}. Equilibrium properties of the $t$--$V$ model have also been studied using neural states on the square lattice~\cite{nys_fermionic_2025} and QMC methods on the honeycomb lattice~\cite{wang_fermionic_2014, wang_stochastic_2016,hesselmann_thermal_2016}.

\section{Microcanonical ensemble}\label{sec:microcanonical}

We start by considering the microcanonical ensemble, which captures the equilibrium properties of a closed system. In ergodic systems, expectation values of local observables in the microcanonical ensemble are typically smooth functions of the energy, as formalized by the eigenstate thermalization hypothesis~\cite{srednicki_approach_1999, rigol_thermalization_2008,dalessio_quantum_2016, deutsch_eigenstate_2018}. The microcanonical ensemble can further be related to the canonical ensemble by ensemble equivalence, which states that canonical expectation values at temperature $T$ are approximately equal to microcanonical expectation values at energy $E = \langle H \rangle_T$. Ensemble equivalence has been established rigorously at high temperature~\cite{kuwahara_gaussian_2020}.

%%%%%%%%%%%%%%%%%%%%%%%%%%%%%%%%%%%%%%%%%%%%%%%%%%%%%%%%%%%%%%%%%%%%%%%%
\subsection{Minimum variance Gaussian states}
\label{sec:min-var-gaussian}
We first try to approximate the microcanonical ensemble by constructing Gaussian states with a small energy variance. To this end, we develop an optimization procedure that may be viewed as gradient descent over the manifold of pure Gaussian states at fixed energy. Parametrizing the Gaussian states in terms of the covariance matrix $\Gamma$, our procedure can be applied to any differentiable, bounded cost function $f(\Gamma)$ subject to differentiable constraints $g_1(\Gamma) = g_2(\Gamma) = \cdots = g_K(\Gamma) = 0$. We define the matrix $F(\Gamma)$ with matrix elements $F_{mn}(\Gamma) = \partial f(\Gamma) / \partial \Gamma_{mn}$ and analogously denote by $G_i(\Gamma)$ the derivative of $g_i(\Gamma)$. Since $\Gamma$ is antisymmetric, the matrices $F$ and $G_i$ can always be chosen to be antisymmetric.

Starting from an initial guess $\Gamma(0)$, we generate a minimization flow $\Gamma(s)$ such that $f(\Gamma(s))$ decreases monotonically, ensuring that the flow reaches a local minimum as $s \to \infty$. Since covariance matrices of pure Gaussian states are related by orthogonal transformations, the flow may be generated by
\begin{equation} \label{eq:flow-generator-B}
    \frac{\di}{\di s} \Gamma(s) = [B(s), \Gamma(s)],
\end{equation}
where $B(s)$ is a real, skew-symmetric matrix. The cost function evolves according to
\begin{equation}
    \frac{\di}{\di s} f(\Gamma) = \Tr \left( F(\Gamma)^T [B, \Gamma] \right) = - \Tr \left( [\Gamma, F(\Gamma)] B \right).
\end{equation}
This expression shows that the choice $B = [\Gamma, F(\Gamma)]^T = [F(\Gamma), \Gamma]$ corresponds to the direction of steepest descent. Taking into account the constraints, we further require that $\di g_i(\Gamma) / \di s = - \Tr ( [\Gamma, G_i(\Gamma)] B) = 0$, which can be realized by projecting $[F(\Gamma), \Gamma]$ onto the subspace orthogonal to $[G_i(\Gamma), \Gamma]$. For multiple constraints, we obtain this projection using a Gram--Schmidt orthogonalization of the constraint directions.

Applied to our problem, we minimize the energy variance $f(\Gamma) = \langle H^2 \rangle_\Gamma - \langle H \rangle_\Gamma^2$ subject to the single constraint $g(\Gamma) = \langle H \rangle_\Gamma - \langle H \rangle_{\Gamma(0)}$, which ensures that the mean energy remains constant during the optimization. We provide an explicit expression for $B$ for a quartic Hamiltonian in Appendix~\ref{appendix-flow-variance}. 

%%%%%%%%%%%%% plot %%%%%%%%%%%%%
\begin{figure}[tbp]
    \centering
    \includegraphics[width=\textwidth]{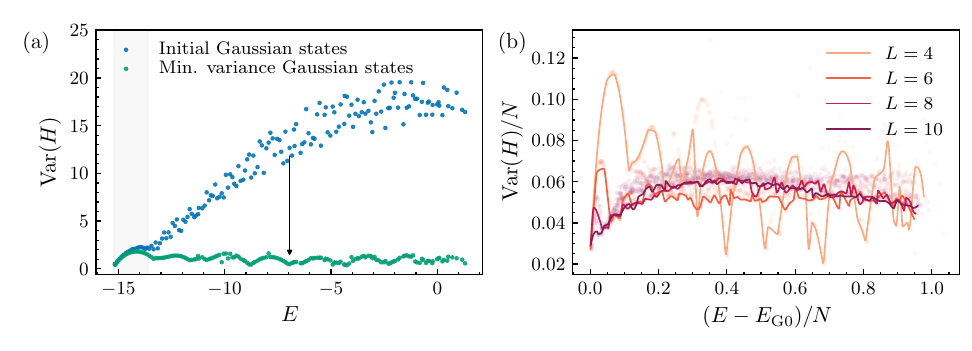}
    \caption{(a)~Energy variance before and after the minimization flow for $V/J=1$ and a lattice of linear size $L=4$ ($N = 16$). (b)~Optimized variance density as a function of the mean energy density for different linear system sizes $L$. The energies are shifted by the lowest energy achieved by a Gaussian state, $E_{G0}$. Each dot represents an individual Gaussian state while the solid line shows the minimum variance among these states within a small energy interval.}
    \label{fig:min-var-L4-system-sizes}
\end{figure}
%%%%%%%%%%%%%%%%%%%%%%%%%%%%%%%%%%%%%%% 

Figure~\ref{fig:min-var-L4-system-sizes} shows the energy variance obtained at the end of the minimization flow as a function of the mean energy of the state. We generate initial states at different energies starting from a random half-filled Gaussian state, which has an expected mean energy equal to $-V N/2(N-1)$. We then lower the energy by evolving the state with imaginary-time TDVP [Eq.~\eqref{eq:imag-time-tdvp}], allowing us to cover the entire range of energies down to the lowest-energy Gaussian state. The imaginary-time evolution preserves particle number and purity. We observe a large reduction of the variance, illustrated in Fig.~\ref{fig:min-var-L4-system-sizes}(a), at high energy. At low energy, the states prepared by TDVP already exhibit a relatively small variance and our minimization procedure only leads to a small improvement. We attribute the ineffectiveness of the minimization to a relative absence of Gaussian states with low energy, consistent with the fact that the TDVP converges to a unique state independent of the initial state. In addition, we observe that the final variance is particularly large in the shaded gray region, which indicates energies below the first excited state, where no eigenstates exist.

In Fig.~\ref{fig:min-var-L4-system-sizes}(b), we explore the dependence of the minimum variance on the system size. We consider square lattices with linear size $L = 4$, $6$, $8$, $10$, corresponding to $N = 16$, $36$, $64$, $100$ fermionic modes. After rescaling the variance and the energy above the lowest energy Gaussian state by $N$, the data appear to converge to a single curve for large systems. This collapse indicates that the minimum achievable variance scales linearly with $N$. Remarkably, the curve is relatively flat across a wide range of energies, contrary to the naive expectation that it might be easier to construct Gaussian states with small energy variance in the middle of the spectrum, where the density of states is high.

The fact that the Gaussian states exhibit $\mathrm{Var}(H) \geq cN$ for some constant $c$ has further implications on the local density of states distribution, which may be of independent interest. We show in Appendix~\ref{appendix-ldos-proof} that the extensive variance implies that the local density of states approaches a Gaussian distribution as $N \rightarrow \infty$. A similar statement, known as the quantum Berry--Esseen theorem~\cite{cramer_berry-esseen_2026}, has been shown to hold for local spin Hamiltonians when the state has exponentially decaying correlations. Our proof applies to arbitrary fermionic Gaussian states, assuming only locality of the interacting Hamiltonian.

%%%%%%%%%%%%% plot %%%%%%%%%%%%%
\begin{figure}[tbp]
    \centering
      \includegraphics[width=\textwidth]{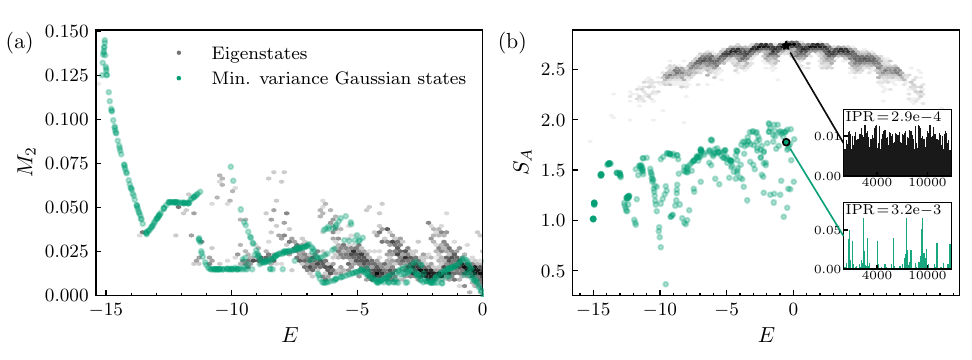}
    \caption{(a)~Order parameter $M_2$ for the minimum-variance Gaussian states and the exact eigenstates of a $4 \times 4$ lattice with $V / J = 1$. (b)~Von Neumann entropy of a $2\times 2$ subsystem for the same states as in (a). The insets show the overlaps of an eigenstate and a minimum-variance Gaussian state at the same mean energy with Fock states in the real-space basis. We also display the corresponding inverse participation ratio (IPR).}
    \label{fig:min-variance-V1-obs-entropy}
\end{figure}
%%%%%%%%%%%%%%%%%%%%%%%%%%%%%%%%%%%%%%%

%%%%%%%%%%%%%%%%%%%%%%%%%%%%%%%%%%%%%%%%%%%%%%%%%%%%%%%%%%%%%%%%%%
Having constructed Gaussian states that minimize the variance at fixed energy, we investigate their properties and, in particular, their resemblance to the microcanonical ensemble. In Fig.~\ref{fig:min-variance-V1-obs-entropy}(a), we compare the order parameter $M_2$ defined in Eq.~\eqref{eq:staggered-density} for the optimized Gaussian states and the exact eigenstates, obtained by exact diagonalization (ED) on a $4 \times 4$ lattice. We observe that the Gaussian states capture some of the extremal values of the eigenstates. However, despite the large number of local minima shown, their expectation values do not densely populate the region in which most of the exact eigenstate values are concentrated.
% This is not unexpected, since restricting the optimization to the Gaussian manifold prevents the resulting states from reproducing generic interacting eigenstates. 

To understand the reason for this poor agreement, we plot the von Neumann entropy of the reduced state on a $2 \times 2$ square in Fig.~\ref{fig:min-variance-V1-obs-entropy}(b). As expected, the eigenstates fall on a smooth curve peaked at the center of the spectrum, with oscillatory features that we attribute to the finite size effects. The subsystem entropy of the minimum-variance Gaussian states is well below this curve. Moreover, the wavefunction of the Gaussian states is concentrated on much fewer Fock states in real space, as shown in the insets of Fig.~\ref{fig:min-variance-V1-obs-entropy}(b). This indicates that the Gaussian states fail to capture the entanglement structure of the eigenstates and thus do not accurately represent the microcanonical ensemble.

%%%%%%%%%%%%%%%%%%%%%%%%%%%%%%%%%%%%%%%%%%%%%%%%%%%%%%%%%%%%%%%%%%%%%%%%
\subsection{Linear combination of Gaussian states from TDVP}
\label{section-beyond-gaussian}
Given the inadequacy of individual Gaussian states to approximate the microcanonical ensemble, we next consider linear combinations of the form $\sum_j c_j |\Psi_G^j\rangle$, where each term $|\Psi_G^j\rangle$ is a Gaussian state. Since overlaps between the different Gaussian states and matrix elements of observables can be computed efficiently with standard techniques (see e.g.~\cite{bravyi_complexity_2016}), the ansatz remains classically tractable for hundreds or even thousands of terms in the sum. The key question is how to choose both the coefficients and states. Most previous approaches parametrize both $c_j$ and $|\Psi_G^j\rangle$ and optimize them according to a desired cost function~\cite{fukutome_theory_1988, ikawa_orbital_1993, koch_linear_1993, jimenez-hoyos_multi-component_2013, snyman_efficient_2021, boutin_quantum_2021, giuliani_precise_2026}. However, this optimization can be computationally demanding because the number of free parameters in each covariance matrix scales as $O(N^2)$ for a general Gaussian state, which limits in practice the number of terms that can be included in the superposition. 

To overcome this challenge, we propose an ansatz based on Krylov-subspace diagonalization techniques, which are widely used to approximate eigenstates of a Hamiltonian~\cite{motta-krylov-nature,motta-krylov-sample-based,motta-krylov-thermal}. 

Following Ref.~\cite{motta-krylov-nature}, we construct the Krylov space using successive applications of the real-time evolution operator $U(\Delta t) = e^{-iH \Delta t}$,which generates the set of states
$\{ \ket{\Psi},\allowbreak U(\Delta t)\ket{\Psi},\allowbreak
U(\Delta t)^2\ket{\Psi},\allowbreak \ldots \}$. For an interacting Hamiltonian, however, the evolved states are in general not Gaussian. To be able to express the basis states efficiently, we replace each time-evolved state by its Gaussian approximation $ U(\Delta t)^k \ket{\Psi} \approx \ket{\Psi_G^k}$, where we denote from now on by $\ket{\Psi_G^k} = \ket{\Psi_G (k \Delta t)}$ the Gaussian state at time $ k \Delta t$  obtained through real-time TDVP evolution as given by Eq.~\eqref{eq:real-time-tdvp}, with $\ket{\Psi_G(0)} = \ket{\Psi}$. Instead of the exact Krylov subspace we hence work with
\begin{equation} \label{eq:krylov-subspace-tdvp}
    V_\Psi = \mathrm{span}\{ \ket{\Psi_G^k} \}_{k=0}^{k_{\max}},
\end{equation}
where $k_{\max}$ determines the dimension of the subspace, assuming that the states are linearly independent, and sets the maximum TDVP evolution time, $t_{\max} = k_{\max} \, \Delta t$.  We note that each state $\ket{\Psi_G^k}$ has the same mean energy since the mean energy is invariant under real-time evolution with TDVP. The subspace $V_{\Psi}$ is therefore a natural choice for estimating observables in the microcanonical ensemble. We denote the projector onto $V_\Psi$ by $\Pi_\psi$.

Given $V_\Psi$, our goal is to find a state in this space with a small energy variance. To this end, we compute the eigenstate of $K_\Psi :=\Pi_\Psi (H-\langle H \rangle_\Psi)^2 \Pi_\Psi$ with the smallest eigenvalue, which we denote by $\ket{\mu_{\Psi}}$. This is indeed equivalent to minimizing $\bra{\psi} K_\Psi \ket{\psi}
= \operatorname{Var}_H(\psi)
+ \bigl(\langle H \rangle_\psi - \langle H \rangle_\Psi)^2$ within the subspace $V_\Psi$, which penalizes both
the energy variance and deviations of the mean energy from that
of the initial state. In practice, we obtain $\ket{\mu_\Psi}$ by solving the generalized eigenvalue problem
\begin{equation}
  K_\Psi \bm c=\lambda S\bm c,
  \qquad
  (K_\Psi)_{ij}=\bra{\Psi_G^i} (H - \langle H \rangle_\Psi)^2 \ket{\Psi_G^j},\quad
  S_{ij}=\bra{{\Psi}_G^i} {\Psi}_G^j \rangle.
  \label{eq:generalized-eigenvalue}
\end{equation}
Given the eigenvector $\bm c$ belonging to the smallest eigenvalue, we obtain $\ket{\mu_\Psi} = \sum_i c_i \ket{\Psi_G^i}$. We highlight that the states generated by TDVP serve only as basis vectors. Even a poor approximation to the exact time evolution may be useful in $V_\Psi$ and late-time snapshots can improve the result.

  %%%%%%%%%%%%%%%%%%%%%%%%%%%%%%%%%%%%%%%%%%%%%%%%%%
 \begin{figure}
 	\centering
    \includegraphics[width=\textwidth]{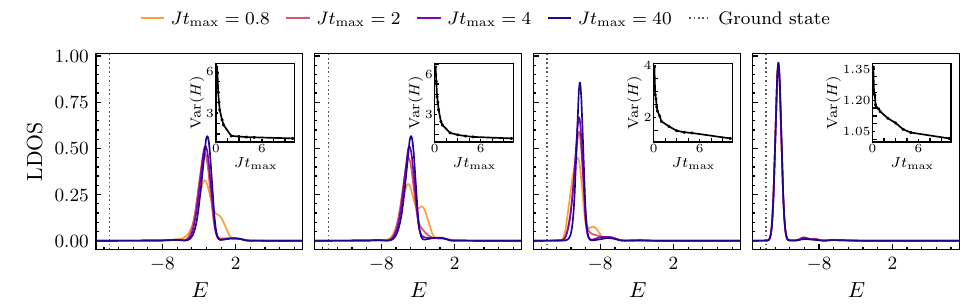}
     \caption{LDOS of $\ket{\mu_\Psi}$ for different initial states $\ket{\Psi}$ across the energy spectrum, for $L = 4$ and $V/J = 1$. The LDOS is computed from the overlaps of $\ket{\mu_\Psi}$ with the exact eigenstates, obtained through ED. To smoothen the discreteness of the spectrum, we apply a Gaussian broadening with standard deviation $\delta=0.4 J$. The different colors indicate different choices of $t_{\max}$, which controls the dimension of the Krylov subspace, for a fixed $J \Delta t = 0.05$. We can see that increasing $t_{\max}$ narrows the spectral distribution of the states. The insets show the energy variance as a function of $t_{\max}$.}
        \label{fig:ldos-energy-methods}
 \end{figure}
  %%%%%%%%%%%%%%%%%%%%%%%%%%%%%%%%%%%%%%%%%%%%%%%%%%

To explore this method numerically, we generate Gaussian states across the energy spectrum by sampling states at different times along the TDVP imaginary-time evolution, following the same procedure as in Sec.~\ref{sec:min-var-gaussian}. From these states, we construct $V_\Psi$~\eqref{eq:krylov-subspace-tdvp} with a time step $\Delta t \propto 1/N$, where $N$ is the system size, up to a maximum time $t_{\max}$. Since the states $\Psi_G^i$  are non-orthogonal, we can retain fewer of them by starting to construct the subspace and adding only states with a significant orthogonal component. This allows to evolve to longer times $t_{\max}$ while keeping a feasible subspace dimension. Figure~\ref{fig:ldos-energy-methods} shows the resulting local density of states (LDOS) for the state $\ket{\mu_\Psi}$ corresponding to different initial states $\ket{\Psi}$ across the energy spectrum, and for different choices of $t_{\max}$. The insets further show that increasing $t_{\max}$ causes the spectral distribution to become narrower. We observe that the variance decreases more slowly with $t_{\max}$ for states at lower energy. This is expected since the fixed point of the imaginary time evolution does not evolve under real-time TDVP such that the subspace $V_\Psi$ corresponding to the lowest-energy Gaussian state is one dimensional.

   %%%%%%%%%%%%%%%%%%%%%%%%%%%%%%%%%%%%%%%%%%%%%%%%%%
 \begin{figure}
 	\centering
    \includegraphics[width=\textwidth]{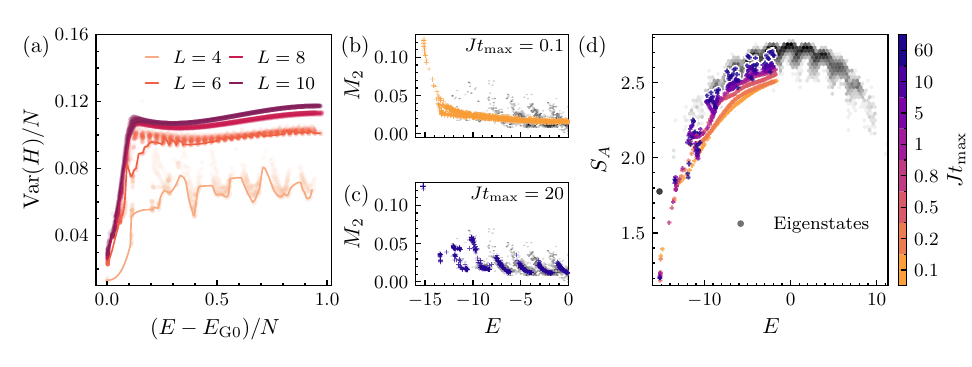}
     \caption{(a)~Variance density of states $\ket{\mu_\Psi}$ plotted as a function of their energy density for $J t_{\max} = 20$. The time steps are $J \Delta t = 0.05, 0.02, 0.0125, 0.008$ for system sizes $L = 4,6,8,10$ ($N = 16,36,64,100$), respectively. (b)~Value of $M_2$ for different states $\ket{\mu_\Psi}$ across all energies for a $4 \times 4$ system with $J t_{\max} = 0.1$. Eigenstates are shown in gray. (c)~Same as (b) with $J t_{\max} = 20$.
     (d)~Von Neumann entropy of the reduced state of the eigenstates and $\ket{\mu_\Psi}$ on a $2 \times 2$ subsystem for different values of $t_{\max}$. The interaction strength is $V / J = 1$ in all plots.}
        \label{fig:krylov-figure-combined}
 \end{figure}
  %%%%%%%%%%%%%%%%%%%%%%%%%%%%%%%%%%%%%%%%%%%%%%%%%%

We next examine the system-size dependence in Fig.~\ref{fig:krylov-figure-combined}(a). We choose a common $t_{\max}$. We observe that the variance $\mathrm{Var}(H)$ remains extensive through the considered energy range although the smaller systems have slightly lower variance per site. As shown in Appendix~\ref{appendix:energy-methods}, the variance could be decreased further by increasing the value of $t_{\max}$. To compare the states $\ket{\mu_\Psi}$ to the exact eigenstates, we again compute the order parameter $M_2$ and the von Neumann entropy of a reduced system. Figures~\ref{fig:krylov-figure-combined}(b) and (c) show that the value of $M_2$ agrees poorly when $t_{\max}$ is small. The agreement is significantly better for larger $t_{\max}$, for which the data points increasingly concentrate in the high-density region of the eigenstate distribution. Similarly, we observe good agreement for the subsystem entropy as well when $t_{\max}$ is sufficiently large, as shown in Fig.~\ref{fig:krylov-figure-combined}(d). These results represent a significant improvement over the corresponding data for the minimum-variance Gaussian states in Fig.~\ref{fig:min-variance-V1-obs-entropy} despite the two sets of states having comparable energy variances. Our findings indicate that the microcanonical ensemble of the $t$--$V$ model exhibits significant non-Gaussian correlations, which can, however, be captured by a linear combination of Gaussians in a subspace constructed from real-time TDVP.

%%%%%%%%%%%%%%%%%%%%%%%%%%%%%%%%%%%%%%%%%%%%%%%%%%%%%%%%%%%%%%%%%%%%%%%%
\section{Canonical ensemble}\label{sec:canonical}
\begin{figure}[t]
    \centering
    \includegraphics[width=\textwidth]{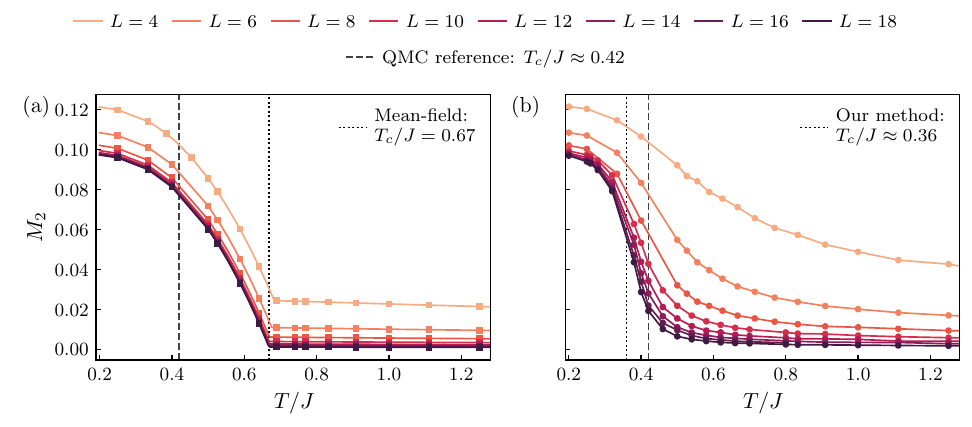}
    \caption{Order parameter $M_2$ as a function of temperature using (a)~a mean-field calculation involving a single Gaussian state, (b)~sampling over Gaussian states according to Eq.~\eqref{monte-carlo-IT-sampling}. We consider system sizes up to $L = 18$ ($N = 324$) with $V / J = 1$. The dashed lines correspond to the critical temperature $T_c$ from Ref.~\cite{gubernatis_two-dimensional_1985,wang_entanglement_2025} while the dotted lines indicate the critical temperature estimated from the respective methods.}
    \label{fig:iGaSampling-vs-gaussian-system-sizes}
\end{figure}

We now proceed to estimating properties in the canonical ensemble at a given inverse temperature $\beta = 1/T$. Using an ansatz of a single Gaussian state, we can obtain an approximation to the Gibbs state $\rho_\beta = e^{-\beta H}/\tr{e^{-\beta H}}$ by minimizing the free energy within the Gaussian manifold. It can be shown that this leads to the self-consistent equation $\Gamma = i\tanh \left[2i\beta h(\Gamma)\right]$ for the covariance matrix $\Gamma$, where $h(\Gamma)$ is the mean-field Hamiltonian defined in Eq~\eqref{eq:real-time-tdvp}. This equation can be solved by fixed-point iteration and provides the standard Gaussian mean-field approximation ~\cite{kraus_generalized_2010,kraus_thesis_fermions,shi_variational_2018}. In Fig~\ref{fig:iGaSampling-vs-gaussian-system-sizes}(a), we plot the resulting value of $M_2$ as a function of the temperature for lattices up to size $18 \times 18$. The mean-field calculation predicts a sharp phase transition at a critical temperature $T_c \approx 0.67 J$, which is substantially different from the transition temperature $T_c \approx 0.42 J$ obtained in QMC studies~\cite{gubernatis_two-dimensional_1985}.

To improve over the mean-field result, we again extend the ansatz beyond a single Gaussian to a linear combination of Gaussian states. As for the microcanonical ensemble, the challenge is to efficiently construct such a linear combination without having to solve a large non-convex optimization problem.
Our approach is based on writing the canonical expectation value $A(\beta) =  {\tr{A e^{-\beta H}}} / {\tr{e^{-\beta H}}}$ as
 \begin{equation} \label{eq:obs-thermal-trace}
 	A(\beta)  = \frac{\sum_\Psi \langle \Psi | e^{-\beta H/2} A e^{-\beta H/2} | \Psi\rangle }{\sum_\Psi  \langle \Psi | e^{-\beta H} | \Psi \rangle },
 \end{equation}
where the sum runs over any complete set of states $\{ \ket{\Psi} \}$ satisfying $\sum_\Psi \ket{\Psi} \bra{\Psi} = \mathds{1}$. By defining
\begin{align} \label{eq:metropolis-IT}
   p_\Psi = \frac{\langle \Psi | e^{-\beta H} | \Psi \rangle}{ \sum_{\Psi'}  \langle {\Psi'} | e^{-\beta H} | \Psi' \rangle}, \qquad  A_\Psi = \frac{\langle \Psi | e^{-\beta H/2} A e^{-\beta H/2} | \Psi\rangle}{ \langle \Psi | e^{-\beta H} | \Psi \rangle},
\end{align}
we may write $A(\beta) = \sum_\Psi p_\Psi A_\Psi$. Using ideas from stochastic  classical methods~\cite{white_minimally_2009,stoudenmire_minimally_2010} and quantum algorithms~\cite{motta-krylov-nature, lu_algorithms_2021}, we may estimate this sum by a Metropolis--Hastings algorithm, where we sample states $\ket{\Psi}$ according to the probability $p_\Psi$. The estimator $\hat{A}(\beta) = \sum_j A_{\Psi_j} / S$ computed with $S$ samples $\ket{\Psi_j}$ satisfies $\mathbb{E}[\hat{A}(\beta)] = A(\beta)$. The algorithm requires the computation of ratios of probabilities $p_{\Psi}/p_{\Psi'} = \frac{\langle \Psi | e^{-\beta H} | \Psi \rangle}{\langle \Psi' | e^{-\beta H} | \Psi' \rangle}$ as well as $A_\Psi$.

%%%%%%%%%%%%%%%%%%%%%%%%%%%%%%%%%%%%%%%%%%%%%%%%%

%%%%%%%%%%%%%%%%%%%%%%%%%%%%%%%%%%%%%%%%%%%%%%%%%

Computing $\pi_\Psi = \langle \Psi | e^{-\beta H} | \Psi \rangle $ and $A_\Psi$ exactly requires performing imaginary time evolution, which is unfeasible in general. Here, we replace the imaginary time evolution in~\eqref{eq:metropolis-IT} by its TDVP approximation, defining $\ket{\Psi_G(\beta)}$ as the Gaussian state obtained by evolving the initial state $\ket{\Psi}$ according to the imaginary-time TDVP in Eq.~\eqref{eq:imag-time-tdvp} for the time $\tau = \beta$. We thus approximate the quantities needed for the Metropolis--Hastings algorithm by
\begin{equation} \label{monte-carlo-IT-sampling}
    \pi_\Psi \approx \Vert \ket{\Psi_G(\beta/2)} \Vert^2, \qquad A_\Psi \approx \frac{\bra{\Psi_G(\beta/2)} A \ket{\Psi_G(\beta/2)}} {\Vert  \ket{\Psi_G(\beta/2)} \Vert^2} .
\end{equation}
Using this approach, we compute the observable $M_2$ for a wide range of temperatures and different system sizes and compare the results obtained to the Gaussian result in Fig.~\ref{fig:iGaSampling-vs-gaussian-system-sizes}. For the numerical experiments presented here, we restrict to the half-filling sector. We take as initial state a Fock state at half-filling, which can be done by choosing $N/2$ occupied modes; we then use as update rule moving a fermion from an occupied to an empty site, which preserves particle number. We initialize independent Markov chains from random Fock states with different degrees of order, including checkerboard configurations and random Fock states, and check equilibration by verifying that the chains produce consistent averages of the observable. After discarding burn-in, we combine of the order of $10^6$ samples per inverse temperature and system size across all the different chains. In comparison to the mean-field result, the curves are much smoother indicating the presence of fluctuations not captured by mean field.

%%%%%%%%%%%%%%%%%%%%%%%%%%%%%%%%%%%%%%%%%%%%%%%%%
\begin{figure}[h]
    \centering
    \includegraphics[width=\textwidth]{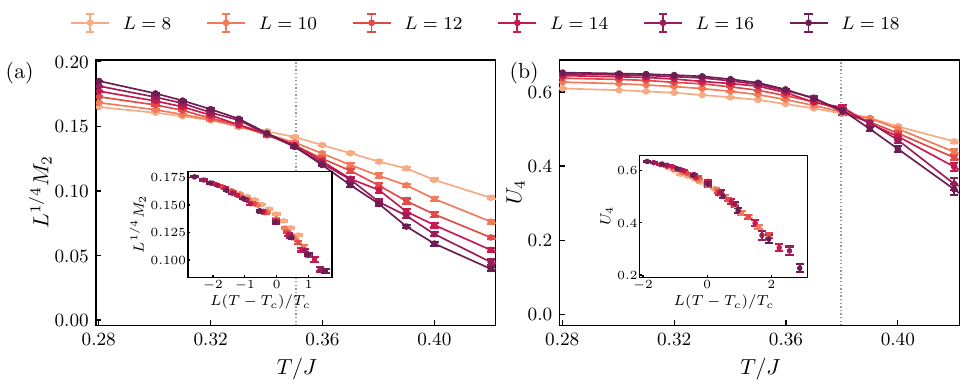}
    \caption{ (a) Scaled observable $L^{1/4} M_2$ and (b) Binder cumulant $U_4$ as a function of temperature for different system sizes ($V / J = 1$), for the sampling algorithm. The insets show the finite-size scaling collapse obtained by rescaling the temperature distance from the critical point as $L^{1/\nu}(T-T_c)/T_c$ using the 2D Ising exponents $\nu=1$ and $\beta_M=1/8$~\cite{hesselmann_thermal_2016,schuckert_many-body_2019}. The vertical lines correspond to the estimated critical temperature.}
    \label{fig:crossing-V1}
\end{figure}

%%%%%%%%%%%%%%%%%%%%%%%%%%%%%%%%%%%%%%%%%%%%%%%%%

We estimate the critical temperature $T_c$ using finite-size scaling arguments for the square magnetization $M_2$ and the Binder cumulant $U_4=1-\langle m^4\rangle/(3M_2^2)$, where $m=\frac{1}{N}\sum_i\eta_i n_i$, as in Eq.~\eqref{eq:staggered-density}. We rescale the square magnetization by $L^{2 \beta_M / \nu}$, where $\beta_M$ and $\nu$ are, respectively, the order-parameter and correlation-length critical exponents. As the phase transition of the $t$--$V$ model lies in the universality class of the 2D Ising model~\cite{gubernatis_two-dimensional_1985, wang_entanglement_2025}, we fix $\beta_M = 1/8$ and $\nu = 1$, leading to a rescaling by $L^{1/4}$. We expect that the rescaled magnetization and the Binder cumulant converge to a finite value in the thermodynamic limit, corresponding to a single crossing point of the curves in Fig.~\ref{fig:crossing-V1}. For the system sizes accessible in our numerical calculations, we still observe a small shift of the crossing temperatures with system size. From the crossing of the $L = 16$ and $L = 18$ data, we obtain $T_c / J =0.353\pm0.008 $ from $L^{1/4} M_2$ and $T_c/J=0.374\pm0.007$ from the Binder cumulant, leading us to the estimate  $T_c/J=0.36\pm0.02$. The estimates obtained from $L^{1/4} M_2$ and $U_4$ approach one another with increasing system size, as also observed in~\cite{hesselmann_thermal_2016}.

This estimated critical temperature is much closer to the QMC estimate $T_c / J = 0.42 \pm 0.04$~\cite{gubernatis_two-dimensional_1985, wang_entanglement_2025} than the mean-field prediction $T_c/J = 0.67$, which we obtained analogously (see Appendix~\ref{appendix:canonical}, which also includes data for $V / J = 2$). Crucially, our approach recovers the correct critical exponents as indicated by the finite-size scaling collapse shown by the insets in Fig.~\ref{fig:crossing-V1}. This is in stark contrast to the mean-field approach, which yields $\beta_M = \nu = 1/2$~\cite{sandvik_computational_2010,cardy_scaling_1996}. Our approach thus provides a qualitative and quantitative improvement over mean-field theory at a modest increase of the computational cost. The remaining discrepancy between our result and the QMC estimate is not unexpected given the approximate nature of imaginary-time TDVP. We note that the QMC results in the literature carry uncertainty, with significant disagreement between different works for $V/J = 2$ (see Appendix~\ref{appendix:canonical}).

\section{Discussion}\label{sec:discussion}
Our results show that individual Gaussian states can only poorly approximate the microcanonical and canonical ensembles of the $t$--$V$ model. In the case of the microcanonical ensemble, we were able to significantly improve the approximation using a linear combination of Gaussian states in a subspace generated by real-time TDVP. As discussed in detail in Appendix~\ref{appendix:energy-methods}, the achievable energy variance continues to decrease for very long evolution times. Future work may explore the potential improvement of these results by, for instance, including longer time evolutions while retaining fewer samples or by enriching the Krylov subspace with different states. Given the variational nature of the approach, the error of real-time TDVP does not directly translate into a final error in the approximation of the microcanonical state. By contrast, our method for the canonical ensemble, which samples from states evolved by imaginary-time TDVP, is ultimately limited by the error of this procedure. Despite this, we observe a significant improvement of the approach over mean-field theory that uses a single Gaussian state to approximate the Gibbs state. 

%%%%%%%%%%%%%%%%%%%%%% plot %%%%%%%%%%%%%%%%%%%%%%%%%%%
\begin{figure}[!htbp]
    \centering
     \includegraphics[width=0.7\textwidth]{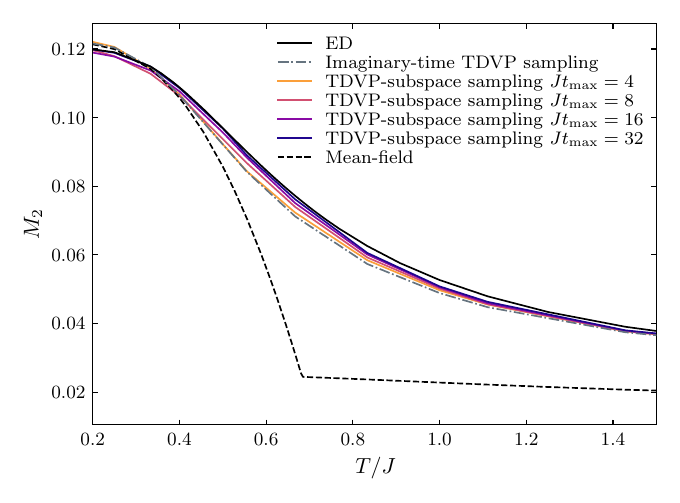}
    \caption{Observable $M_2$ as a function of temperature for four different methods: (i) Exact diagonalization (ED), (ii) imaginary-time TDVP sampling method of Section~\ref{sec:canonical}, (iii) subspace method introduced in Section~\ref{sec:discussion}, and (iv) mean-field theory. Method (iii) uses a time step $J \Delta t = 0.05$ and we show four different maximum evolution times $J t_{\max}$. Results were obtained for a $4 \times 4$ system with $V/J = 1$.}
    \label{fig:methods-temperature-vs-ED}
\end{figure}
%%%%%%%%%%%%%%%%%%%%%%%%%%%%%%%%%%%%%%%%%%%%%%%%%

It is possible to combine the sampling approach for the canonical ensemble with the variational subspace used to approximate the microcanonical ensemble. For any subspace $V_{\Psi}$ with projector $\Pi_\Psi$, we can define the projected Hamiltonian $\tilde{H}_\Psi = \Pi_{\Psi} H \Pi_{\Psi}$ and replace Eq.~\eqref{eq:metropolis-IT} by 
\begin{equation} \label{eq:monte-carlo-krylov-sampling}
    p_\Psi = \frac{\langle \Psi | e^{-\beta \tilde{H}_\Psi} |\Psi\rangle}{\sum_{\Psi'} \langle \Psi' | e^{-\beta \tilde{H}_{\Psi'}} |\Psi' \rangle}, \qquad A_\Psi = \frac{\langle \Psi| e^{-\beta \tilde{H}_\Psi/2} A e^{-\beta \tilde{H}_\Psi/2}|\Psi\rangle}{\langle \Psi| e^{-\beta \tilde{H}_\Psi} |\Psi\rangle} .
\end{equation}
Note that $e^{-\beta \tilde{H}_\Psi/2}|\Psi\rangle$ is, in general, a linear superposition of Gaussian states, whereas the state $\ket{\Psi_G(\beta/2)}$ in Eq.~\eqref{monte-carlo-IT-sampling} is a single Gaussian state. This ansatz is therefore, in principle, more expressive. To test this idea, we implement this approach with the variational subspace defined in Eq.~\eqref{eq:krylov-subspace-tdvp}. In Fig.~\ref{fig:methods-temperature-vs-ED}, we compare the  observable values obtained using this new approach with the previous approach of Eq.~\eqref{monte-carlo-IT-sampling}. For large values of $t_\mathrm{max}$, the subspace method moves closer to the exact result, indicating that subspace methods can combat the error incurred by imaginary-time TDVP. We note that the subspace method is computationally more expensive as each sample requires the computation of matrix elements in the subspace. The method is nevertheless a promising candidate for large-scale numerical studies, which may be systematically improved by enlarging the subspace. Future work may, for instance, include states from both real- and imaginary-time TDVP in the subspace $V_\Psi$. It will be particularly interesting to explore the performance of these methods in more strongly correlated systems such as the Fermi--Hubbard model.

%%%%%%%%%%%%%%%%%%%%%%%%%%%%%%%%%%%%%%%%%%%%%%%%%
%%%%%%%%%%%%%%%%%%%%%%%%%%%%%%%%%%%%%%%%%%%%%%%%%
%%%%%%%%%%%%%%%%%%%%%%%%%%%%%%%%%%%%%%%%%%%%%%%%%

% \section{Conclusion}
\begin{acknowledgments}
We thank Tao Shi and \'Alvaro M. Alhambra for insightful discussions. The work is supported by the German Federal Ministry of Education and Research (BMBF) through the funded project ALMANAQC, grant number 13N17236 within the research program ``Quantum Systems''. The research is partly funded by THEQUCO as part of the Munich Quantum Valley, which is supported by the Bavarian state government with funds from the Hightech Agenda Bayern. E.C. used the OpenAI models Sol and Astra to assist with specific derivations, manuscript polishing and improving the efficiency of specific numerical experiments. The authors take full responsibility for the content of this manuscript.
\end{acknowledgments}

%%%%%%%%%%%%%%%%%%%%%%%%%%%%%%%%%%%%%%%%%%%%%%%%%
\clearpage
\appendix
\section{Minimization flow for the variance of quartic Hamiltonians}
\label{appendix-flow-variance}
In this Appendix, we provide the derivation of the minimization flow introduced in~\ref{sec:min-var-gaussian} for the particular case of minimizing the variance at fixed mean energy. As defined in the main text, let $c_i$ be Majorana operators with $\{c_i,c_j\}=2\delta_{ij}$, $T$ a real antisymmetric matrix, and $U$ a real fully antisymmetric tensor. Summation over repeated indices is understood. We express the quartic Hamiltonian $H$ as $H = H_2 + H_4$, with
\begin{equation}
H_2=i T_{ij}c_i c_j,
\qquad
H_4=U_{ijkl}c_i c_j c_k c_l. 
\end{equation}
Let $\rho$ be a pure Gaussian state and define its covariance matrix $\Gamma$ by $\Gamma_{ij} = i (\langle c_i c_j\rangle_\rho - \delta_{ij})$, with $\Gamma^T=-\Gamma$ and $\Gamma^2=-I$. Define the mean-field matrix and its commutator with the covariance by
\begin{equation}
{h_{ij}=T_{ij}-6U_{ijkl}\Gamma_{kl}},
\qquad A=2[h,\Gamma].
\end{equation}
Here $h$ corresponds to the mean-field Hamiltonian defined in the main text in Eq.~\eqref{eq:real-time-tdvp}.
We first show that the variance $f(\Gamma)=\langle H^2\rangle_\rho-\langle H\rangle_\rho^2$ can be written as
\begin{equation} \label{eq:appendix-variance}
\begin{aligned}
 f(\Gamma)&=\frac14\operatorname{tr}(A^T A)+24U_{ijkl}U_{mnpq}\bigl(
 \Gamma_{im}\Gamma_{jn}\Gamma_{kp}\Gamma_{lq}\\
 &\hspace{12mm}-6\delta_{im}\delta_{jn}\Gamma_{kp}\Gamma_{lq}
 +\delta_{im}\delta_{jn}\delta_{kp}\delta_{lq}\bigr).
 \end{aligned}	
\end{equation}
To prove this, we apply Wick's theorem. The mean energy $E=\langle H\rangle_\rho$ is 
\begin{equation}
{E=T_{ij}\Gamma_{ij}-3U_{ijkl}\Gamma_{ij}\Gamma_{kl}.}
\end{equation}
The variance can be decomposed into connected Wick contributions from $H_2^2$, $H_4^2$, and the cross terms $\{H_2, H_4\}$. {Pairings with no contractions between the two factors cancel against $\langle H\rangle_\rho^2$.} $H_2^2$ contributes the term
\begin{equation}
	 2T_{ij}T_{mn} \operatorname{Re}\bigl[({\delta_{im}-i\Gamma_{im}})
                        ({\delta_{jn}-i\Gamma_{jn}})\bigr]  =2T_{ij}T_{mn}(\delta_{im}\delta_{jn}-\Gamma_{im}\Gamma_{jn}),
\end{equation}
whereas $H_4^2$ contributes the two different terms
\begin{align}
& 72U_{ijkl}U_{mnpq}\Gamma_{kl}\Gamma_{pq}(\delta_{im}\delta_{jn}-\Gamma_{im}\Gamma_{jn}), \\
& 24U_{ijkl}U_{mnpq}\operatorname{Re}\bigl[({\delta_{im}-i\Gamma_{im}})({\delta_{jn}-i\Gamma_{jn}})({\delta_{kp}-i\Gamma_{kp}})({\delta_{lq}-i\Gamma_{lq}})\bigr],
\end{align}
corresponding to Wick pairings connecting each of the quartic terms by two and four pairs, respectively. Finally, the cross terms contribute
\begin{equation}
	{-24T_{ij}U_{mnpq}\Gamma_{pq}(\delta_{im}\delta_{jn}-\Gamma_{im}\Gamma_{jn}).}
\end{equation}
The antisymmetry of $T$ and $U$ gives the corresponding prefactors. The quadratic contribution, the two-pair quartic contribution, and the cross contribution combine into
\begin{equation}
2h_{ij}h_{mn}(\delta_{im}\delta_{jn}-\Gamma_{im}\Gamma_{jn})
=-2\operatorname{tr}(h\Gamma h\Gamma)-2\operatorname{tr}(h^2)
=\frac14\operatorname{tr}(A^T A),
\end{equation}
where the last equality uses $\Gamma^2=-I$. Adding the four-pair quartic contribution establishes Eq.~\eqref{eq:appendix-variance}.
%%%%%%%%%%%%%%%%%%%%%%%%%%%%%%%%%%%%%%%%%%%%%%%%%

We next derive the expression for the operator $B$ introduced in Eq.~\eqref{eq:flow-generator-B} that defines the minimization flow for the variance at fixed mean energy. Take $\dot\Gamma=[B,\Gamma]$ with $B^T=-B$.  

\paragraph*{Derivative of the variance.}
Let us take the derivative of $f(\Gamma)$ as defined in Eq.~\eqref{eq:appendix-variance}. For the first term, note that $\dot A=2[\dot h,\Gamma]+2[h,\dot\Gamma]$ and ${\dot h_{ij}=-6U_{ijkl}\dot\Gamma_{kl}}$. Differentiating $\Gamma^2=-I$ gives $\{\Gamma,\dot\Gamma\}=0$. Define
\begin{equation}
{(N_1)_{ij}=4(h\Gamma h)_{ij}-24U_{ijkl}(\Gamma h\Gamma+h)_{kl}.}
\end{equation}
Using the cyclicity of the trace and this tangent condition gives

\begin{align}
\frac{d}{dt}\left[\frac14\operatorname{tr}(A^T A)\right]
&=-\frac12\operatorname{tr}(A\dot A)\notag\\
&=-4\operatorname{tr}\bigl((\Gamma h\Gamma+h)\dot h\bigr)
  -2\operatorname{tr}\bigl((2h\Gamma h-\{h^2,\Gamma\})\dot\Gamma\bigr) = -\operatorname{tr}(N_1\dot\Gamma). \notag
\end{align}
For the remaining terms in Eq.~\eqref{eq:appendix-variance}, antisymmetry
of $U$ makes the four derivatives of the quartic product equal, and
likewise the two derivatives of the quadratic product. Define
\begin{align}
(N_2)_{im}&=96U_{ijkl}U_{mnpq}\Gamma_{jn}\Gamma_{kp}\Gamma_{lq},\\
(N_3)_{kp}&=-288U_{ijkl}U_{mnpq}\delta_{im}\delta_{jn}\Gamma_{lq}.
\end{align}
Their contributions are then $-\operatorname{tr}((N_2+N_3)\dot\Gamma)$; the constant term has zero derivative. All three matrices are
antisymmetric; with $N=N_1+N_2+N_3$, we obtain
\begin{equation}
\dot f=-\operatorname{tr}(N\dot\Gamma)
      =\operatorname{tr}([N,\Gamma]B).
\end{equation}

\paragraph*{Energy constraint and flow.}
Differentiating the mean energy given above, using the antisymmetry properties of $T,U$ and $\Gamma$ gives

\begin{equation}
\dot E=(T_{ij}-6U_{ijkl}\Gamma_{kl})\dot\Gamma_{ij}
      =-\operatorname{tr}(h\dot\Gamma)
      =\operatorname{tr}([h,\Gamma]B).
\end{equation}
{In the notation of the main text, the variance derivative satisfies $[F,\Gamma]=[N,\Gamma]$, while the energy constraint $g(\Gamma)=E(\Gamma)-E(\Gamma(0))$ has derivative matrix $G=h$.}
Let $\|\cdot\|_F$ denote the Frobenius norm. Without the energy constraint, taking $B=[N,\Gamma]$ gives
$\dot f=-\|[N,\Gamma]\|_F^2\leq0$. To impose $\dot E=0$, we need
to take $B$ orthogonal to $[h,\Gamma]$ in the Frobenius inner product. For $[h,\Gamma]\neq0$, we subtract the orthogonal projection of $[N,\Gamma]$ onto $[h,\Gamma]$, giving:
\begin{equation}
B=[N,\Gamma]
-\frac{\operatorname{tr}([N,\Gamma][h,\Gamma])}
       {\operatorname{tr}([h,\Gamma]^2)}[h,\Gamma].
\end{equation}
The denominator is nonzero since $\operatorname{tr}([h,\Gamma]^2)=-\|[h,\Gamma]\|_F^2$.
To check that the constraint is satisfied, note that
$\operatorname{tr}([h,\Gamma]B)=0$, so that $\dot E=0$. Similarly,
using the antisymmetry of $B$, we have:
\begin{equation}
\dot f=\operatorname{tr}([N,\Gamma]B)=\operatorname{tr}(B^2)
      =-\|B\|_F^2\leq0.
\end{equation}
Thus, the evolution
$\dot\Gamma=[B,\Gamma]$ preserves purity and keeps the energy constant,
while decreasing the variance whenever $B\neq0$.

%%%%%%%%%%%%%%%%%%%%%%%%%%%%%%%%%%%%%%%%%%%%%%%%%%%%%%%%%%%%%%%%%%%%%%
%%%%%%%%%%%%%%%%%%%%%%%%%%%%%%%%%%%%%%%%%%%%%%%%%%%%%%%%%%%%%%%%%%%%%%
\clearpage
\section{Analysis of the TDVP-generated Gaussian superposition ansatz}
\label{appendix:energy-methods}
In this Appendix, we investigate more in depth the variance reduction, as well as  the dimension of the subspace defined in Eq.~\eqref{eq:krylov-subspace-tdvp} in the main text, as a function of the evolution time $J t_{\max}$. Since this subspace is constructed with real-time TDVP evolved states, a key question is whether the TDVP trajectory keeps exploring the Hilbert space and adding useful terms to the superposition as we increase the maximum evolution time, and whether this implies further reduction on the variance of the ansatz.
 \begin{figure}[H]
\centering
\includegraphics[width=\linewidth]{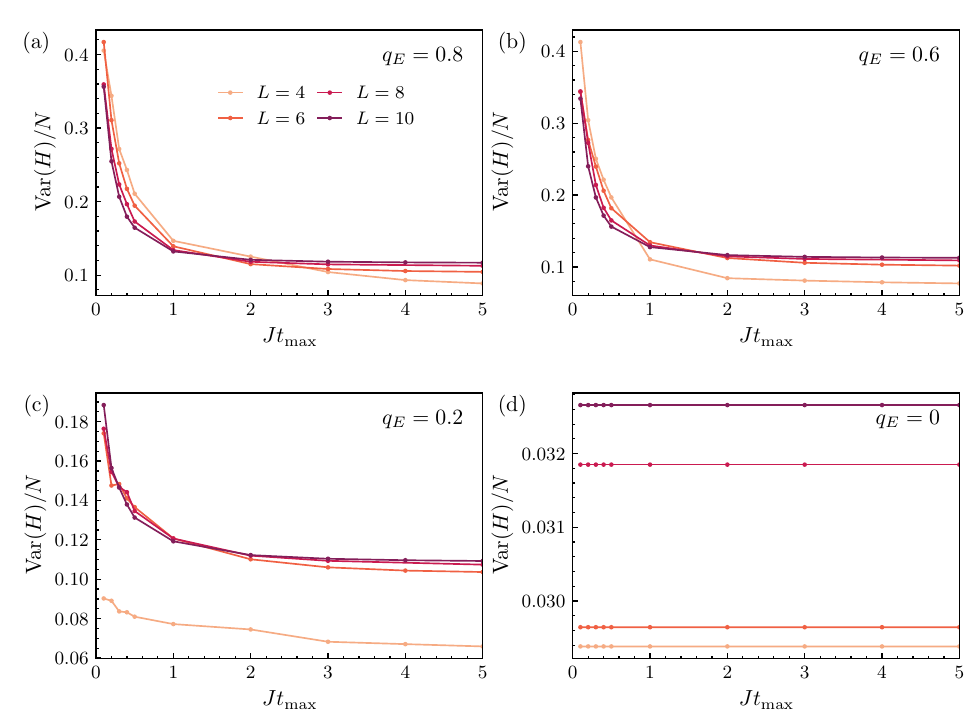}
\caption{Short-time variance per site for $L=4,6,8,10$. The lowest-energy states do not change over time, as expected, since they are stationary states with respect to the TDVP real-time evolution and their corresponding subspace $V_\Psi$ is one-dimensional. Here $V/J = 1$.}
\label{fig:short-time}
\end{figure}
To address this, we first show in Fig.~\ref{fig:short-time} the variance density as a function of $J t_{\max}$, for different system sizes $L = 4, 6, 8, 10$ (corresponding to $N = 16, 36, 64, 100$), for four initial states at comparable energy density. We define $q_E = (\langle H\rangle - E_{G_0})/(E_{\max}^{\mathrm{samp}} - E_{G_0})$, where $E_{G_0}$ is the lowest energy obtained with Gaussian states and $E_{\max}^{\mathrm{samp}}$ is the maximum energy we have obtained in our samples, which corresponds on average to $-VN/2(N-1)$.  What we observe is that the variance rapidly decreases as we increase $J t_{\max}$, and we reach an apparent plateau where this reduction slows down. 

\begin{figure}[H]
\centering
\includegraphics[width=\linewidth]{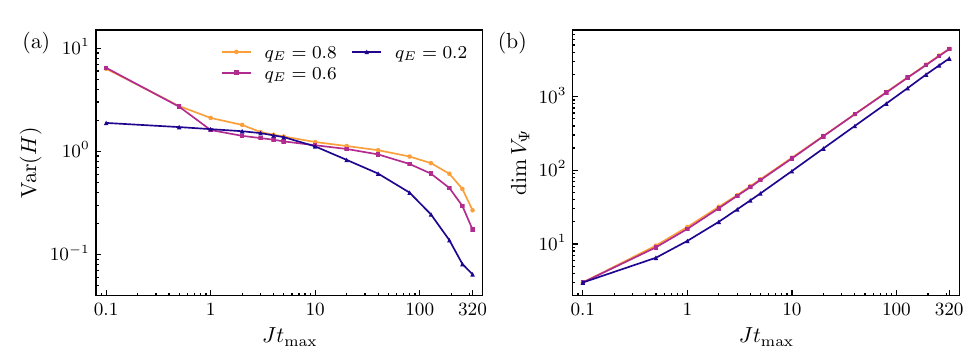}
\caption{$L=4$: (a) energy variance and (b) dimension of the subspace $V_\Psi$. Here $V/J = 1$.}
\label{fig:l4-long}
\end{figure}

To understand better whether we can keep decreasing the variance by increasing $Jt_{\max}$, we explore next what happens for longer times for $L = 4$,  a system size for which we can reach long evolution times without much computational effort. Fig.~\ref{fig:l4-long} extends the maximum evolution times up to $Jt_{\max}=320$, for fixed $J\Delta t=0.05$. We plot in Fig.~\ref{fig:l4-long}(a) the variance across different times. We observe that it keeps decreasing: even though it reaches an apparent plateau at intermediate times, it drops for late times and reaches values of the order $10^{-1}$. Fig.~\ref{fig:l4-long}(b) shows the dimension of the subspace $V_\Psi$, which keeps increasing as we evolve with longer times. To exploit our construction while saving computation memory and time, we build the  subspace by adding Gaussian states from the TDVP trajectory when they contain a sufficiently large component outside the span of the states already selected. In practice, a candidate $|\Psi_G^j\rangle$ is retained if $\|(I-\Pi_\Psi)|\Psi_G^j\rangle\|^2>\eta_{\mathrm{tol}}$, where $\Pi_\Psi$ denotes as usual the projector into the current subspace. We choose in practice $\eta_{\mathrm{tol}}=10^{-10}$ in these examples. This avoids adding states to the basis that are highly non-orthogonal with the ones that are already in place.

\begin{figure}[htbp]
\centering
\includegraphics[width=\linewidth]{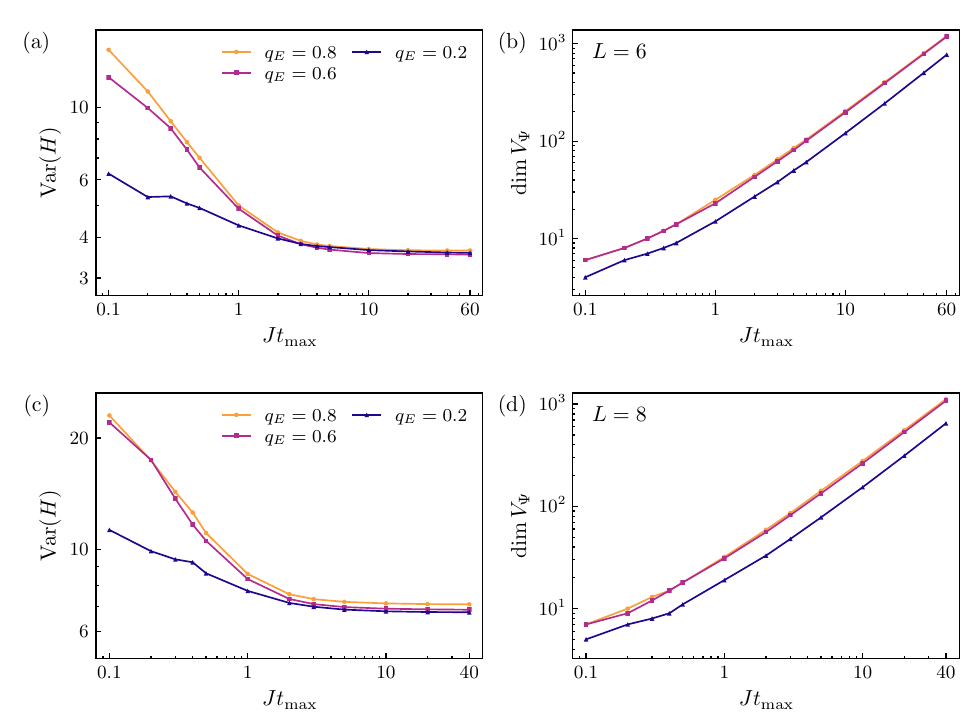}
\caption{ (a) Energy variance  and (b) dimension of the subspace $V_\Psi$, for $L = 6$. (c) and (d) are the corresponding ones for $L = 8$. Here $V/J = 1$.}
\label{fig:larger-sizes}
\end{figure}

In Fig.~\ref{fig:larger-sizes} we represent analogous plots for  $L = 6$ and $L = 8$. We observe that the dimension of the subspace continues to grow  as $t_{\max}$ increases, showing that the TDVP trajectory continues to generate new directions in the Hilbert space. We want to emphasize that our ansatz uses of the order of hundreds to thousands of Gaussian states in its superposition. As for the variance, while it keeps decreasing, the improvement slows down. Whether we can observe a further decrease and a departure from the apparent plateau, as observed for $L = 4$, in accessible evolution times remains open. Alternative ways of enriching the subspace could provide further improvements here, which we leave for future exploration.

%%%%%%%%%%%%%%%%%%%%%%%%%%%%%%%%%%%%%%%%%%%%%%%%%%%%%%%%%%%%%%%%%%%%%%
%%%%%%%%%%%%%%%%%%%%%%%%%%%%%%%%%%%%%%%%%%%%%%%%%%%%%%%%%%%%%%%%%%%%%%
\clearpage
\section{Additional results for canonical ensemble}
\label{appendix:canonical}
\begin{figure}[h]
    \centering
    \includegraphics[width=\textwidth]{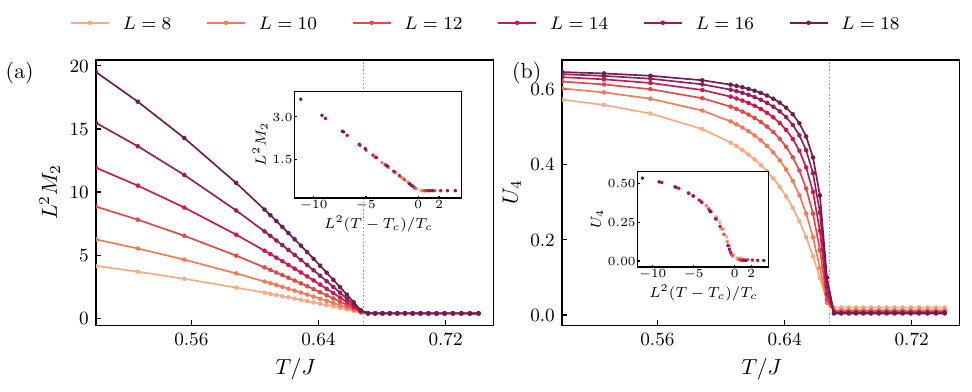}
    \caption{(a) Scaled observable $L^{2} M_2$ and (b) Binder cumulant $U_4$ as a function of temperature for different system sizes ($V / J = 1$) for the mean-field solution. The insets show the finite-size scaling collapse obtained by rescaling the temperature distance from the critical point as $L^{1/\nu}(T-T_c)/T_c$ using the mean-field exponents $\beta_M = \nu=1/2$~\cite{sandvik_computational_2010}. The dotted vertical lines indicate the temperature at which the curves for the two largest system sizes cross.}
    \label{fig:appendix-canonical-mean-field-crossing-V1}
\end{figure}

In this Appendix, we provide complementary data to the canonical ensemble results presented in Section~\ref{sec:canonical}. In Fig.~\ref{fig:appendix-canonical-mean-field-crossing-V1} we present the results for the scaled observable $L^2 M_2$ and the Binder cumulant $U_4$ for the mean-field results, which serve to provide the estimation of the critical temperature given in the main text. The rescaling of the square magnetization by $L^2$ corresponds to the mean-field critical exponents $\beta_M = \nu = 1/2$. As before, the critical temperature is estimated from the crossing for the biggest system sizes, $L = 16$ and $L = 18$, of both the observable and Binder cumulant. We obtain $T/J = 0.671$ and $T/J = 0.668$, respectively, which we combine to $T_c/J \approx 0.67$.

\begin{figure}[h]
    \centering
    \includegraphics[width=\textwidth]{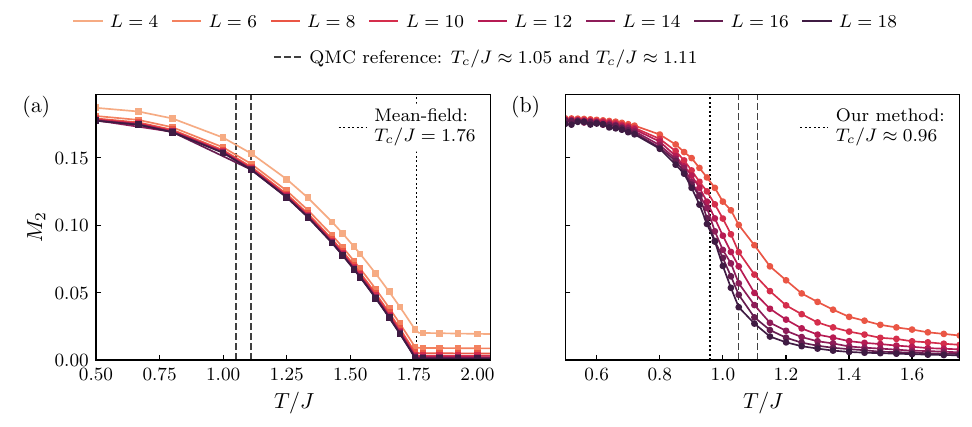}
    \caption{Order parameter $M_2$ as a function of temperature using (a)~a mean-field calculation involving a single Gaussian state, (b)~sampling over Gaussian states according to Eq.~\eqref{monte-carlo-IT-sampling}. We consider system sizes up to $L = 18$ with $V / J = 2$. The dashed lines correspond to the critical temperatures $T_c$ from Refs.~\cite{gubernatis_two-dimensional_1985,wang_entanglement_2025}.}
    \label{fig:appendix-canonical-compare-V2}
\end{figure}

\begin{figure}[h]
    \centering
    \includegraphics[width=\textwidth]{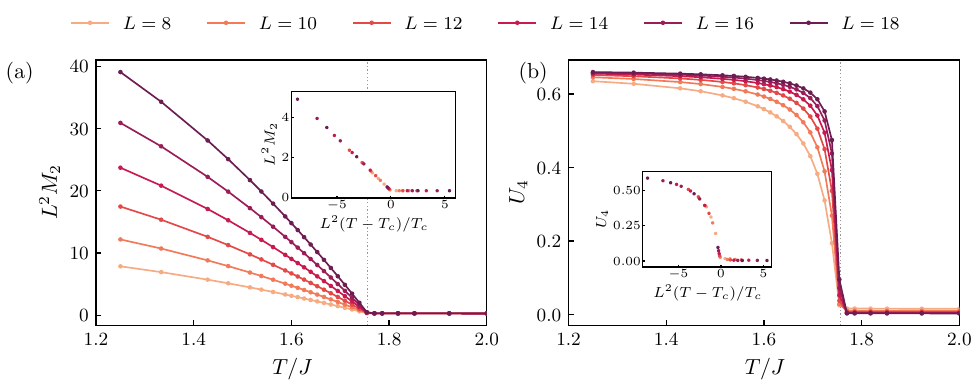}
    \caption{(a)~Scaled observable $L^{2} M_2$ and (b)~Binder cumulant $U_4$ as a function of temperature for different system sizes ($V / J = 2$) for the mean-field solution. The insets show the finite-size scaling collapse obtained by rescaling the temperature distance from the critical point as $L^{1/\nu}(T-T_c)/T_c$ using the mean-field exponents $\beta_M = \nu=1/2$~\cite{sandvik_computational_2010}. The dotted vertical lines indicate the temperature at which the curves for the two largest system sizes cross.}
    \label{fig:appendix-canonical-mean-field-crossing-V2}
\end{figure}

\begin{figure}[h]
    \centering
    \includegraphics[width=\textwidth]{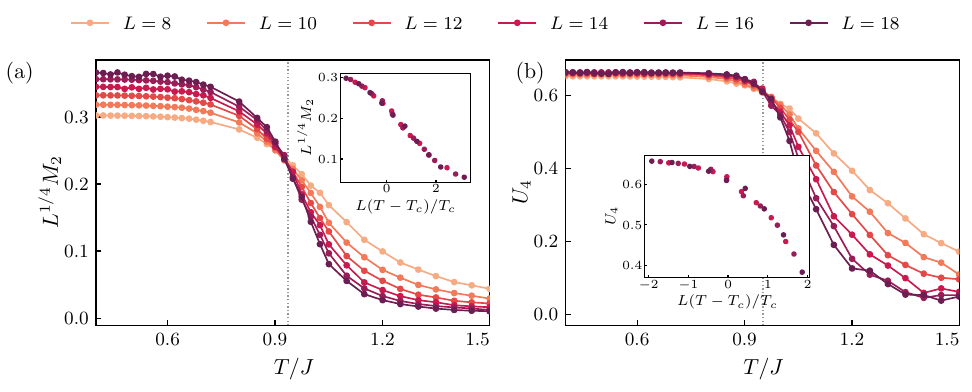}
    \caption{(a) Scaled observable $L^{1/4} M_2$ and (b) Binder cumulant $U_4$ as a function of temperature for different system sizes ($V / J = 2$), for the sampling algorithm. The insets show the finite-size scaling collapse obtained by rescaling the temperature distance from the critical point as $L^{1/\nu}(T-T_c)/T_c$ using the 2D Ising exponents $\nu=1$ and $\eta=1/4$~\cite{hesselmann_thermal_2016,schuckert_many-body_2019}. The dotted vertical lines indicate the temperature at which the curves for the two largest system sizes cross.}
    \label{fig:appendix-canonical-method-crossing-V2}
\end{figure}

Fig.~\ref{fig:appendix-canonical-compare-V2} shows analogous results to Fig.~\ref{fig:iGaSampling-vs-gaussian-system-sizes} in the main text for $V/J = 2$. The estimates of the critical temperature are obtained again from the crossings of the scaled observable ($L^2 M_2$ for mean field and $L^{1/4} M_2$ for our sampling method) and the Binder cumulant $U_4$ for the biggest system sizes, $L=16$ and $L=18$. These crossings are presented in Fig.~\ref{fig:appendix-canonical-mean-field-crossing-V2} and Fig.~\ref{fig:appendix-canonical-method-crossing-V2} for the mean-field approach and our sampling method, respectively. For the mean-field calculation, these crossings occur at $T/J=1.756$ for $L^2 M_2$ and $T/J=1.769$ for $U_4$, respectively. For our sampling method, the corresponding crossings occur at $T/J=0.958$ and $T/J=0.952$.

%%%%%%%%%%%%%%%%%%%%%%%%%%%%%%%%%%%%%%%%%%%%%%%%%%%%%%%%%%%%%%%%%%%%%%
%%%%%%%%%%%%%%%%%%%%%%%%%%%%%%%%%%%%%%%%%%%%%%%%%%%%%%%%%%%%%%%%%%%%%%
\clearpage
\section{Local density of states of fermionic Gaussian states}
\label{appendix-ldos-proof}
In this Appendix, we prove that the local density of states (LDOS) of a fermionic Gaussian state approaches a Gaussian distribution in the thermodynamic limit, for local Hamiltonians with quadratic and quartic interactions, provided that the energy variance of the state grows at least linearly with the system size. The main ingredient is a uniform bound on the energy cumulants, whose absolute values are at most $O(N)$ at each fixed order. This bound also provides a bound on the convergence rate.

Related results have been shown for quantum spin systems with local interactions. For product states, expectation values of observables with disjoint supports factorize, which allows to prove central limit theorems for the energy distribution~\cite{hartmann_gaussian_2004}. More recently, Berry--Esseen bounds have provided improved convergence rate estimates, both for product states as well as for states with exponentially decaying correlations~\cite{rai_matrix_2024, cramer_berry-esseen_2026}.

This Gaussian approximation of the LDOS has numerous applications. It is a central ingredient in proofs of equivalence between the microcanonical and canonical ensembles~\cite{brandao_equivalence_2015}, in the study of thermalization of quantum many-body systems \cite{farrelly_thermalization_2017}, and in proving the efficiency of energy-filtering algorithms~\cite{rai_matrix_2024}. We refer the reader to~\cite{cramer_berry-esseen_2026} for an overview of these applications and further references.

Here, we establish a corresponding result for fermionic Gaussian states without imposing spatial decay of correlations. Such states need not factorize over disjoint supports and can exhibit long-range correlations. Instead, Wick’s theorem expresses their higher-order correlations through two-point functions, which allow us to control their energy cumulants and show the convergence of the LDOS. We obtain an $O(N^{-1/6})$ bound on the convergence rate between the cumulative LDOS and the cumulative Gaussian distribution. Whether this rate can be improved under the same assumptions remains open.

We proceed now to state the main assumptions and results. Let $\rho$ be a fermionic Gaussian state on $N$ fermionic modes, and consider the Hamiltonian
\begin{equation} \label{eq:H-definition}
	H = i \sum_{pq} W^{(2)}_{pq}\gamma_p \gamma_q + \sum_{pqrs}  W^{(4)}_{pqrs} \gamma_p \gamma_q  \gamma_r  \gamma_s 
\end{equation}
where $\gamma_p = \gamma_p^\dagger$ denote Majorana operators, with $\{ \gamma_p, \gamma_q \} = 2 \delta_{pq}$ and $p = 1,\ldots, 2N$. $W^{(2)}$ is a real and antisymmetric tensor, corresponding to the quadratic part. $W^{(4)}$ is a real and completely antisymmetric tensor, corresponding to the quartic part. We assume bounded Hamiltonian terms as well as sparsity of the tensors.  Formally, we impose that $|W^{(2)}_{pq} | \leq J_2$ and $|W^{(4)}_{pqrs}| \leq J_4$, for all indices $p,q,r,s$, with $J_l$, $l = 2,4$, a constant independent of the system size $N$. Besides, we impose the following sparsity assumption: for each fixed index $p$, we have that there are at most $D_2$ non-zero entries $W^{(2)}_{pq}$, and at most $D_4$ non-zero entries $W^{(4)}_{pqrs}$, with $D_l$ independent of the system size, $l = 2,4$.  These two conditions together imply the following: 
\begin{align} \label{eq:bound-tensors-entries}
	\max_p \sum_{q} |W^{(2)}_{pq} | \leq J_2 D_2, \qquad \max_p \sum_{qrs} |W^{(4)}_{pqrs} | \leq J_4 D_4.
\end{align}
Note that these conditions are quite general and include models such as the $t$--$V$ and Fermi--Hubbard models. We note that it is possible to extend the assumptions to include more general local Hamiltonians with uniformly bounded terms of constant support, provided that each Majorana operator appears in only a constant number of terms. For such Hamiltonians, bounds analogous to Eq.~\eqref{eq:bound-tensors-entries} can be derived.

The LDOS and its characteristic function are
\begin{equation}
 D(E)=\tr{\rho\delta(H-E)},\qquad
 \phi(t)=\int dE\,D(E)e^{\ii Et}
        =\langle e^{\ii tH}\rangle_\rho.
 \label{eq:ldos_def}
\end{equation}
We define the energy cumulants $\kappa_m(H)$ through the derivatives of the logarithm of $\phi(t)$ as~\cite{doring_method_2021}:
\begin{equation}
 \kappa_m(H)
 :=(-\ii)^m
   \left.\frac{\di^m}{\di t^m}\log\phi(t)\right|_{t=0},
 \qquad m\geq1.
 \label{eq:cumulant_expansion}
\end{equation}
We note that this is well-defined since $\phi(t)$ is analytic for any finite $N$ and satisfies $\phi(0)=1$ such that the $\log \phi(t)$ is also analytic at $t = 0$.
We observe that $\kappa_1(H)=\langle H\rangle_\rho$ is the mean energy and $\kappa_2(H)=\langle H^2\rangle_\rho-\langle H\rangle_\rho^2$ is the energy variance. We first establish the following bound.

\begin{theorem}[$O(N)$ bound on all cumulants]
\label{thm:energy_cumulants}
Let $H_N$ be a Hamiltonian defined on $N$ fermionic modes as in Eq.~\eqref{eq:H-definition} and assume Eq.~\eqref{eq:bound-tensors-entries} holds with constants independent of $N$. Let $\rho_N$ denote a fermionic Gaussian state. Then, there exists a constant $K=K(J_2,J_4,D_2,D_4)$, independent of $m$, $N$, and the state $\rho_N$, such that every fermionic Gaussian state satisfies
\begin{equation}
 |\kappa_m(H)|\leq N(m!)^2K^m,\qquad m\geq2.
 \label{eq:main_cumulant_bound}
\end{equation}
\end{theorem}

In particular, the energy variance is at most $O(N)$. If the variance of $\rho_N$ is also bounded below by $cN$, with $c>0$ independent of $N$, then for every fixed $m\geq3$ the normalized cumulant $\kappa_m(H)/\kappa_2(H)^{m/2}$ tends to zero as $N\rightarrow \infty$. Since a Gaussian distribution has vanishing cumulants beyond second order, this gives convergence to a Gaussian distribution. The bound on the cumulants also gives the following convergence rate result. 

\begin{corollary}[Local density of states of Gaussian states]\label{cor:LDOS_gaussian} 
Let $H_N$ denote a Hamiltonian defined on $N$ fermionic modes that satisfies Eq.~\eqref{eq:bound-tensors-entries}, with constants independent of $N$. Let $\rho_N$ denote a sequence of fermionic Gaussian states. We denote the LDOS of $\rho_N$ by $D_N(E)$, its mean energy by $\mu_N=\langle H_N\rangle_{\rho_N}$, and its energy variance by $\sigma_N^2=\kappa_2(H_N)$. Assume, in addition, that
\begin{equation}
 \sigma_N^2\geq cN,
 \label{eq:variance_lower_bound}
\end{equation}
where $c>0$ is independent of $N$. Define
\begin{equation}
 F_N(E)=\int_{-\infty}^E dE'\,D_N(E'),\qquad
 G_N(E)=\int_{-\infty}^E dE'\,\mathcal N(E';\mu_N,\sigma_N),
\end{equation}
where $\mathcal N(E;\mu,\sigma)=(2\pi\sigma^2)^{-1/2}\exp[-(E-\mu)^2/(2\sigma^2)]$. Then
\begin{equation}
 \zeta_N:=\sup_{E\in\mathbb R}\abs{F_N(E)-G_N(E)}
 \leq C_{\mathrm{LDOS}}N^{-1/6},
 \label{eq:zeta_LDOS_stat}
\end{equation}
where $C_{\mathrm{LDOS}}$ depends only on the Hamiltonian parameters and $c$, and is independent of $N$ and the sequence of states.
\end{corollary}

\subsection{Proof of Thm.~\ref{thm:energy_cumulants}}
To prove Thm.~\ref{thm:energy_cumulants}, we first establish norm bounds on the Hamiltonian tensors based on Eq.~\eqref{eq:bound-tensors-entries}. Wick's theorem then expresses each cumulant $\kappa_m$ as a sum over diagrams with $m$ interaction vertices, while the cumulant definition cancels all disconnected contributions. Each resulting diagram can be seen as a tensor network contraction, which we can bound in terms of the norms of its tensors. To obtain the desired bound, we first transform the network into an equivalent one by absorbing problematic vertices while controlling the norms of the resulting tensors. This procedure constitutes the most technical part of the proof and gives an $O(N)$ bound for each connected diagram. Finally, counting the Wick pairings gives the claimed cumulant bound.

%%%%%%%%%%%%%%%%%%%%%%%%%%%%%%%%%%%%%%%%%%%%%%%%%%%%%%%%
\subsubsection{Definitions and norm bounds}
Given a matrix $M$, we recall its operator norm defined by 
\begin{equation}
 \norm{M}_{\mathrm{op}}
 :=\sup_{x\neq0}\frac{\norm{Mx}_{\ell^2}}{\norm{x}_{\ell^2}}
 \label{eq:matrix_norm_definitions}
\end{equation}
where $\ell^2$ is the usual vector 2-norm. For a $k$-index tensor $T$ with elements $T_{i_1, \ldots, i_k}$, we define a matricization $T^{(A|B)}$ as the matrix obtained by grouping its indices into two sets $A$ and $B$, which form the row and column indices, respectively. $\norm{T^{(A|B)}}_{\mathrm{op}}$ is then the ordinary operator norm in Eq.~\eqref{eq:matrix_norm_definitions}. If either set $A$ or $B$ is empty, the matricization is a row or column vector and the corresponding operator norm is simply the $\ell^2$ vector norm, which is equal to the Frobenius norm
\begin{equation}
    \| T \|_F = \sqrt{\sum_{x_1, \ldots, x_k} |T_{x_1, \ldots, x_k}|^2}.
\end{equation}
The Frobenius norm satisfies
\begin{equation}
    \| T \|_F \leq \sqrt{2 N} \| T^{(A | B)} \|_\op, \qquad A = \{1\}, \quad B = \{2, \ldots k\},
    \label{eq:f_norm}
\end{equation}
which follows from the fact that the Frobenius norm is the square root of the sum of squared singular values of $T^{(A|B})$ while the operator norm is equal to the largest singular value. The same expression holds for any matricization in which $A$ or $B$ contains a single index.

The operator norm satisfies the bound
\begin{equation} \label{ineq-2-1-infty-norm}
	\Vert M \Vert_\op \leq \sqrt{\Vert M \Vert_1 \Vert M \Vert_\infty},
\end{equation}
where $\| \cdot \|_1$ and $\| \cdot \|_\infty$ are the matrix norms induced by the $\ell^1$ and $\ell^\infty$ vector norms. In particular, $\Vert M \Vert_1 = \max_j \sum_{i} |M_{ij}|$ and $\Vert M \Vert_\infty := \max_i \sum_{j}|M_{ij}|$. For Hamiltonian tensors as in Eq.~\eqref{eq:H-definition} and Eq.~\eqref{eq:bound-tensors-entries}, we set 
\begin{equation}
	\Lambda:=\max\{J_2D_2,\,J_4D_4\}.
\end{equation}
Antisymmetry makes Eq.~\eqref{eq:bound-tensors-entries} valid with any one index fixed. Fixing more indices only decreases the sums, so every nontrivial matricization has absolute row and column sums at most $\Lambda$. Therefore Eq.~\eqref{eq:bound-tensors-entries} and the matrix inequality~\eqref{ineq-2-1-infty-norm} imply for $l \in \{2, 4\}$
\begin{align}
 \label{eq:ab_norm}
 \|W^{(l),(A|B)}\|_\op &\leq \Lambda, \qquad A,B \ne \varnothing .
 % \|W^{(l)}\|_F &\leq \sqrt{2N}\,\Lambda,
 % \label{eq:f_norm}
\end{align}
We state a further bound that will become useful when manipulating the tensor network later. For any $2N\times2N$ matrix $R$, define the matrix $T(R)$ in terms of its matrix elements
\begin{equation}
	 T(R)_{cd}=\sum_{a,b}W^{(4)}_{abcd}R_{ab}.
\end{equation}
It holds that
\begin{equation}
 \|T(R)\|_\op\leq\Lambda\|R\|_\op.
 % \|T(R)\|_F\leq\sqrt{2N}\,\Lambda\|R\|_\op.
 \label{eq:absorb}
\end{equation}
To see this, we note that $|R_{ab}|\leq\|R\|_\op$. For each fixed index $c$,
\begin{equation}
	 \sum_d|T(R)_{cd}|
 \leq\|R\|_\op\sum_{a,b,d}|W^{(4)}_{abcd}|
 \leq\Lambda\|R\|_\op.
\end{equation}
The same argument with $d$ fixed bounds the column sums. Equation~\eqref{ineq-2-1-infty-norm} then yields the desired bound on the operator norm.

%% Cov matrix %%

Finally, we will require a bound on the operator norm of the covariance matrix. We recall the definition of the covariance matrix $\Gamma_{p q} = \frac{\ii}{2}\Tr (\rho[\gamma_p,\gamma_q] )$ and define the Hermitian two-point matrix
\begin{equation}
 C_{pq}=\langle\gamma_p\gamma_q\rangle_\rho =\delta_{pq}-\ii\Gamma_{pq}.
 \label{eq:covariance_convention}
\end{equation}
Throughout this section, we will work exclusively with $C$ instead of $\Gamma$. In a slight abuse of terminology, we will also refer to $C$ as the covariance matrix. Physical states satisfy  $\|\Gamma\|_\op\leq1$ and thus
\begin{equation}
    \Vert C \Vert_\op\leq 2.
    \label{eq:C_norms}
\end{equation}

\subsubsection{Cumulants as connected tensor network contractions}
\begin{figure}[htbp]
\centering
\begin{tikzpicture}[
  font=\small,
  graphvertex/.append style={minimum size=5mm}
]
  % Left panel: wider spacing and larger boxes.
  \begin{scope}[shift={(0,0.25)},scale=1.15,transform shape]

    \node[font=\scriptsize,align=center] at (2.55,0.90)
      {each arc $(k,\ell)$ contributes $C_{x_kx_\ell}$};

    \foreach \x/\k in {
      0/1,0.50/2,1.00/3,1.50/4,
      2.30/5,2.80/6,
      3.60/7,4.10/8,4.60/9,5.10/10
    } {
      \fill (\x,0) circle (1.5pt);
      \node[font=\scriptsize,below=3pt]
        at (\x,0) {$x_{\k}$};
    }

    % Groups corresponding to the three insertions.
    \draw[rounded corners,gray!75]
      (-0.25,-0.45) rectangle (1.75,0.55);
    \draw[rounded corners,gray!75]
      (2.05,-0.45) rectangle (3.05,0.55);
    \draw[rounded corners,gray!75]
      (3.35,-0.45) rectangle (5.35,0.55);

    \node at (0.75,-0.70) {$A_1$};
    \node at (2.55,-0.70) {$A_2$};
    \node at (4.35,-0.70) {$A_3$};

    % Pairings between different insertions.
    \draw[treeedge,bend left=31] (1.00,0) to (2.30,0);
    \draw[treeedge,bend left=38] (1.50,0) to (3.60,0);
    \draw[treeedge,bend left=28] (2.80,0) to (4.10,0);

    % More pronounced pairings within each insertion.
    \draw[residual,line width=1pt,
          dash pattern=on 2pt off 1pt]
      (0,0)
      .. controls (0,0.60) and (0.50,0.60) ..
      (0.50,0);

    \draw[residual,line width=1pt,
          dash pattern=on 2pt off 1pt]
      (4.60,0)
      .. controls (4.60,0.60) and (5.10,0.60) ..
      (5.10,0);

  \end{scope}

  % Longer arrow and more space between panels.
  \draw[prune] (6.65,0.30)--(7.85,0.30);

  % Corresponding network.
  \node[graphvertex] (g1) at (8.50,0.40) {$1$};
  \node[graphvertex] (g2) at (9.30,-0.45) {$2$};
  \node[graphvertex] (g3) at (10.10,0.40) {$3$};

  \draw[treeedge] (g1)--(g2)--(g3)--(g1);

  \draw[residual]
    (g1.north west)
    .. controls +(-0.45,0.65) and +(0.45,0.65) ..
    (g1.north east);
  \draw[residual]
    (g3.north west)
    .. controls +(-0.45,0.65) and +(0.45,0.65) ..
    (g3.north east);

  \node[font=\scriptsize,red!75!black]
    at (8.50,1.25) {self-loop};
  \node[font=\scriptsize,red!75!black]
    at (10.10,1.25) {self-loop};

\end{tikzpicture}
\caption{A Wick pairing of three operators $A_1$, $A_2$, $A_3$ (left) and
its corresponding tensor network (right). Pairings within the same operator become self-loops.}
\label{fig:example-Wick-diagram}
\end{figure}
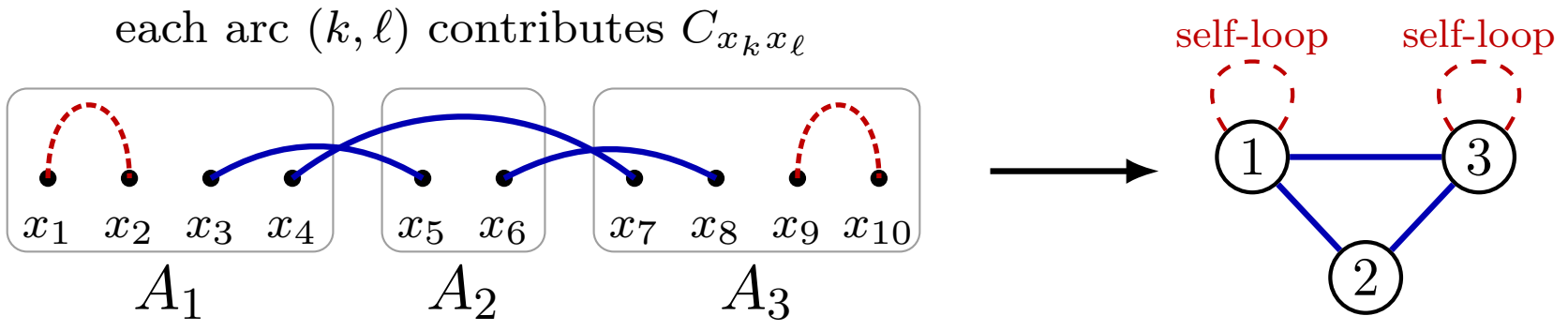

Here we show how to express the cumulants defined in Eq.~\eqref{eq:cumulant_expansion} through Wick's theorem. Note that since the cumulants involve the expectation value of powers of the Hamiltonian, we need to compute expressions like $\langle\gamma_{x_1}\cdots\gamma_{x_P}\rangle_\rho$, which are given by Wick's theorem as a sum over the product of $2$-point strings as
\begin{equation}
 \langle\gamma_{x_1}\cdots\gamma_{x_P}\rangle_\rho
 =\sum_{\pi\in\mathrm{Pair}(P)}\operatorname{sgn}(\pi)
                      \prod_{(k,\ell)\in\pi}C_{x_kx_\ell},\qquad k<\ell.
 \label{eq:wick}
\end{equation}
Here, $\mathrm{Pair}(P)$ denotes all pairings of the $P$ positions in the ordered Majorana string, and $\operatorname{sgn}(\pi)$ denotes the sign of the permutation that brings paired positions together.  Each pairing groups these positions into disjoint pairs, with every position appearing exactly once. Given an operator of the form $A_1 A_2 \cdots A_k$ with $A_\mu = \sum_{i_1, \ldots, i_{l_\mu}} a^{\mu}_{i_1, \ldots, i_{l_\mu}} \gamma_{i_1} \cdots \gamma_{i_{l_\mu}}$, we can express its expectation value as a tensor network contraction as shown in Fig.~\ref{fig:example-Wick-diagram}. Each vertex represents a tensor $a^{\mu}_{i_1, \ldots, i_{l_\mu}}$ while the lines carry a matrix $C$. In our setting, the tensors $a^{\mu}$ will be one of the Hamiltonian tensors $W^{(2)}$ or $W^{(4)}$.

%%%%%%%%%%% Figure %%%%%%%%%%%%
%%%%%%%%%%%%%%%%%%%%%%%

We note that only connected Wick diagrams / tensor networks contribute to the cumulant \cite{chen_convergence_2026} (see also Appendix D in \cite{giuliani_gentle_2021}, or Appendix C in \cite{greenblatt_adiabatic_2024}), a result that is commonly referred to as the \emph{linked cluster theorem } \cite{coleman_introduction_2015,negele_quantum_2018}. 
Consequently, the cumulant may be expressed as a sum of connected tensor networks,
\begin{equation}
 \kappa_m(H)
 =\sum_{\mathcal N\ \mathrm{connected},\ m\ \mathrm{sites}}
   \varepsilon_{\mathcal N}\,\mathcal C(\mathcal N).
 \label{eq:cumulant_network_sum}
\end{equation}
The sum runs over connected tensor networks with $m$ vertices, where each vertex can be a two-leg or a four-leg tensor corresponding to $W^{(2)}$ and $W^{(4)}$. The coefficient $\varepsilon_{\mathcal N}$ collects the Wick sign as well as the factors $\ii$ from the quadratic contributions. The term $\mathcal{C}(\mathcal{N})$ is the contraction of the tensor network as outlined above. With tensors $W^{(l_1)}, \ldots, W^{(l_m)}$ on sites $1, \ldots, m$ ($l_i \in \{2, 4\}$), we obtain the explicit expression
\begin{equation}
    \mathcal C(\mathcal N) = \sum_{x_1, \ldots, x_L} W^{(l_1)}_{x_1\ldots x_{l_1}} W^{(l_2)}_{x_{l_1+1}\ldots x_{l_1+l_2}} \cdots W^{(l_m)}_{x_{L-l_m+1}\ldots x_L}
    \prod_{(i, j)\in E(\mathcal N)} C_{x_i x_j},
    \label{eq:network_value}
\end{equation}
where $L = \sum_{i=1}^m l_i$ and $E(\mathcal{N})$ is the set of edges of the tensor network $\mathcal{N}$. Each $x_i$ runs from $1$ to $2N$ as usual.

\subsubsection{Cumulant bound}
In order to obtain a bound for the cumulant, we first bound the number of terms in the sum of Eq.~\eqref{eq:cumulant_network_sum}. For each vertex, we can choose either $W^{(2)}$ or $W^{(4)}$, resulting in a total of $2^m$ choices. For each choice, there are at most $4m$ tensor indices to be paired up by edges. The number of distinct pairings resulting in a connected tensor network is bounded by the number of all possible pairings, which is at most
\begin{equation}
 (4m-1)!!
 =\prod_{j=1}^m \bigl[(4j-3)(4j-1)\bigr]
 \leq\prod_{j=1}^m 16j^2
 =16^{m} (m!)^2.
 \label{eq:pairing_count_explicit}
\end{equation}

\begin{figure}[htbp]
\centering
\begin{tikzpicture}[font=\small]
  \node[localtensor,minimum width=10mm,minimum height=10mm] (w) at (0.3,0) {$W^{(4)}$};
  \draw[cedge] ($(w.west)+(0,0.22)$)--++(-0.9,0) node[left] {$p$};
  \draw[cedge] ($(w.west)+(0,-0.22)$)--++(-0.9,0) node[left] {$q$};
  \draw[cedge] ($(w.east)+(0,0.22)$)
     .. controls +(0.8,0.65) and +(0.8,-0.65) ..
     ($(w.east)+(0,-0.22)$);
  \node[right=7mm of w] (clabel) {$C_{rs}$};

  \draw[prune] (2.3,0)--(3.5,0)
       node[midway,above=4pt] {};

  \node[localtensor,minimum width=13mm,minimum height=10mm] (b) at (5.0,0)
       {$\widehat W^{(2)}$};
  \draw[cedge] ($(b.west)+(0,0.22)$)--++(-0.65,0);
  \draw[cedge] ($(b.west)+(0,-0.22)$)--++(-0.65,0);
\end{tikzpicture}
\caption{Example of absorption of a self-loop into a new quadratic tensor $\widehat W$. The new tensor retains the same scaling in its norm bounds as $W$.}
\label{fig:self-loop-dressing}
\end{figure}
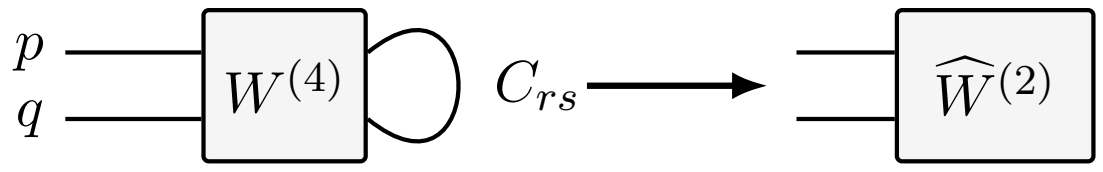

Next, we bound the magnitude $\mathcal C (\mathcal N)$ in terms of the norm bounds of each of the tensors that compose $\mathcal N$. As a first step, we show that self-loops can be removed as shown in Fig.~\ref{fig:self-loop-dressing}. For a connected network with $m\geq2$, such a loop can occur only on a quartic tensor.  A loop on a quadratic tensor or two loops on a quartic tensor would imply a disconnected part of the network, which is not allowed. We may thus remove a single loop by defining
\begin{equation}
 \widehat W^{(2)}_{pq}=\sum_{r,s}W^{(4)}_{prsq}C_{rs}.
 \label{eq:self_loop_dressing}
\end{equation}
Applying Eq.~\eqref{eq:absorb} with $R=C$ together with Eq.~\eqref{eq:C_norms} yields
\begin{equation}
	 \|\widehat W^{(2)}\|_{\op}\leq2\Lambda 
 % \qquad
 % \|\widehat W^{(2)}\|_{\mathrm F}\leq2\sqrt{2N}\,\Lambda.
\end{equation}
Thus removing a self-loop effectively increases the operator norm of a local tensor by a factor of $2$.

Having removed all loops in this manner, we wish to bring the vertices into a so-called $st$-ordering~\cite{even_computing_1976}. Given vertices $s$ and $t$ in a graph with $m$ vertices, an $st$-ordering assigns the number $1$ to $s$ and the number $m$ to $t$. All other vertices are numbered $\{2, \ldots, m-1\}$ in such a way that each vertex is connected by at least one edge to a vertex with a smaller number and by at least one edge to a vertex with a bigger number. Such an ordering may not exist for all graphs. A sufficient condition for a $st$-ordering to exist is that the graph contains no articulation vertex, i.e., a vertex whose removal will disconnect the graph~\cite{even_computing_1976}. Our goal therefore is to first remove all articulation vertices from the tensor network.

\begin{figure}[htbp]
\centering
\includegraphics[width=\linewidth]{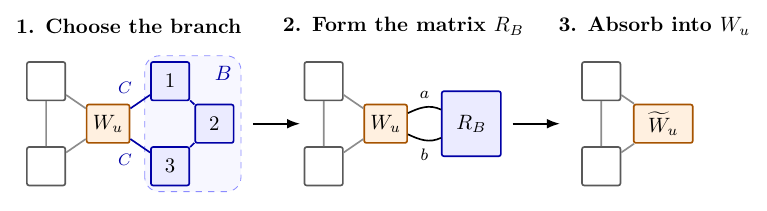}
\caption{Steps to absorb a branch that is connected to the rest of the tensor network by a single vertex $u$. Here, tensors $1,2,3$ are connected to the rest of the network only through the quartic tensor $W_u$. We first contract the branch and both boundary edges, each carrying a covariance matrix $C$, into the matrix $R_B$; we have denoted by $(a,b)$ the corresponding matrix indices. We then absorb $R_B$ into $W_u$ using~\eqref{eq:absorb}, leaving the two legs toward the rest of the network. Since the norm of $R_B$ can be controlled, this procedure does not worsen our bound.}
\label{fig:branch-absorption-steps}
\end{figure}

Suppose removing a tensor at vertex $u$ disconnects the network, as in Fig.~\ref{fig:branch-absorption-steps}. Since all degrees are two or four, each resulting component is attached to $u$ by an even number of edges. Hence $u$ has four incident edges, with exactly two edges to each of two components. Since there may be more than one vertex for which this happens, we choose $u$ such that its removal leaves the smallest branch $B$ disconnected from the rest of the tensor network. Then, the branch $B$ together with $u$ has no articulation vertex. This implies that $B$ either contains a single vertex or that $u$ is connected to two distinct vertices in $B$, which we label by $b_1$ and $b_n$, where $n = |B|$. Moreover, $B$ must be connected since each tensor carries an even number of indices.

Next, we contract $B$, including the two covariance matrices $C$ from the edges connecting $B$ to $u$, into a two-index tensor $R_B$, which we view as a matrix.  We further form a new tensor $\widetilde W_u$ by contracting $W_u$ with $R_B$ as shown in Fig.~\ref{fig:branch-absorption-steps}. By Eq.~\eqref{eq:absorb},
\begin{equation}
     \|\widetilde W_u\|_{\op}\leq\Lambda\|R_B\|_{\op}.
\end{equation}
Suppose $u$ is connected to two distinct vertices in $B$. Since $\{u\} \cup B$ contains no articulation vertices, we can bring $B$ into $st$-ordering, with $s = b_1$ the first vertex and $t = b_n$ the last. Since all tensors have at least one input and one output leg with respect to this ordering, we can express $R_B$ as a product of matrices $E_1 V_1 E_2 V_2 \cdots V_n E_{n+1}$. Up to tensor products with the identity matrix and index permutations, each $V_i$ is a matricization of $W^{(l_i)}$ and $E_i$ is a tensor product of covariance matrices $C$. By using the fact that $\|AB\|_{\mathrm{op}}\leq\|A\|_{\mathrm{op}}\|B\|_{\mathrm{op}}$ together with Eq.~\eqref{eq:C_norms} and Eq.~\eqref{eq:ab_norm}, we obtain
\begin{equation}
    \|R_B\|_{\mathrm{op}}\leq 2^{|E(B)|} \Lambda^{|B|},
\end{equation}
where $E(B)$ is the set of edges in $B$, including the two edges connecting it to $u$. The bound automatically accounts for the potential presence of loops and remains valid for the special case when $B$ consists of a single vertex.

The reduced network again has degrees two or four, so we can repeat this procedure while accounting for the increase in operator norm of some of the tensors in each step. Once no articulation vertices remain, we may find an $st$-ordering of the remaining vertices. Except for the first and last vertex in the ordering, we may apply the operator norm bound as above. After this procedure, each original vertex of the tensor network contributes a factor $\Lambda$ while each edge contributes a factor $2$. The first and last vertex in the ordering require special treatment because the corresponding matrices have no row or no column index, i.e., they are vectors. Their operator norm is therefore equal to the Frobenius norm, which, according to Eq.~\eqref{eq:f_norm} incurs an additional factor $\sqrt{2 N}$. We thus arrive at the bound
\begin{equation}
     |\mathcal C(\mathcal N)|
 \leq2N\Lambda^m2^{|E(\mathcal N)|}
 \leq2N(4\Lambda)^m.
\end{equation}
Combining this with the $2^m$ choices of rank-2 or rank-4 tensors for each site and the pairing count~\eqref{eq:pairing_count_explicit} gives
\begin{equation}
     |\kappa_m(H)|\leq2N(128\Lambda)^m(m!)^2
\end{equation}
which proves Thm.~\ref{thm:energy_cumulants} with $K = 256 \Lambda$.

%%%%%%%%%%%%%%%%%%%%%%%%%%%%%%%%%%%%%%%%%%%%%%%%%%%%%%%%
%%%%%%%%%%%%%%%%%%%%%%%%%%%%%%%%%%%%%%%%%%%%%%%%%%%%%%%%

\subsection{Proof of Corollary.~\ref{cor:LDOS_gaussian}}
We keep the same notation as before, where $\rho_N$ denotes a Gaussian state, and $H_N$ a Hamiltonian under assumptions~\eqref{eq:H-definition}--~\eqref{eq:bound-tensors-entries}. For convenience, choose the energy origin so that $\mathrm{tr}\left( \rho_N H_N \right) = 0$. This can always be achieved by subtracting the mean energy from $H_N$, which leaves all cumulants of order at least two unchanged.

We compare the LDOS with the Gaussian distribution having the same mean and variance, through their cumulative distributions:
\begin{equation}
 F_N(E)=\int_{-\infty}^{E} D_N(E')\,dE',
 \qquad
 G_N(E)=\frac{1}{\sqrt{2\pi}\sigma_N}
        \int_{-\infty}^{E}
        e^{-E'^2/(2\sigma_N^2)}\,dE'.
\end{equation}
After normalizing by the energy standard deviation $\sigma_N$, the  cumulant of order $m$ becomes $\kappa_m(H_N)/\sigma_N^m$. The bound from
Thm.~\ref{thm:energy_cumulants} therefore gives
\begin{equation}
 \frac{|\kappa_m(H_N)|}{\sigma_N^m}
 \leq
 \frac{N(m!)^2K^m}{(cN)^{m/2}}
 =
 (m!)^2\frac{K^2}{c}
 \left(\frac{K}{\sqrt{cN}}\right)^{m-2},
 \qquad m\geq3.
 \label{eq:normalized_cumulant_bound}
\end{equation}
For every fixed $m\geq3$, this tends to zero as $N$ increases.
Thus the higher cumulants of the normalized LDOS approach
their Gaussian values.

To obtain the claimed convergence rate, we use the following consequence of Theorem~2.4 in~\cite{doring_method_2021}. 

\begin{theorem}[Theorem 2.4 in~\cite{doring_method_2021}] If for some $\Delta_N>0$,
\begin{equation}
 \frac{|\kappa_m(H_N)|}{\sigma_N^m}
 \leq \frac{(m!)^2}{\Delta_N^{m-2}},
 \qquad m\geq3,
 \label{eq:statulevicius_ldos_condition}
\end{equation}
then
\begin{equation}
 \sup_{E\in\mathbb R}|F_N(E)-G_N(E)|
 \leq C_0\Delta_N^{-1/3},
 \label{eq:statulevicius_ldos_bound}
\end{equation}
where $C_0$ is a universal constant.
\end{theorem}
To verify~\eqref{eq:statulevicius_ldos_condition}, choose
\begin{equation}
 \Delta_N
 :=\frac{\sqrt{cN}}
 {K\max\{1,K^2/c\}}.
\end{equation}
Since $m-2\geq1$, we have
$\max\{1,K^2/c\}^{m-2}\geq K^2/c$, so~\eqref{eq:normalized_cumulant_bound} implies~\eqref{eq:statulevicius_ldos_condition}. Moreover, $\Delta_N$ is proportional to $\sqrt N$. Consequently,
\begin{equation}
 \zeta_N
 =\sup_{E\in\mathbb R}|F_N(E)-G_N(E)|
 \leq C_{\mathrm{LDOS}} N^{-1/6},
\end{equation}
where $C_{\mathrm{LDOS}}$ depends only on $K$ and $c$. This proves
Corollary.~\ref{cor:LDOS_gaussian}.
%%%%%%%%%%%%%%%%%%%%%%%%%%%%%%%%%%%%%%%%%%%%%%%%%%%%%%%%
%%%%%%%%%%%%%%%%%%%%%%%%%%%%%%%%%%%%%%%%%%%%%%%%%%%%%%%%

\bibliography{beyond_gaussian_paper_clean}
\end{document}